\documentclass[12pt]{article}
\usepackage{epsfig,rotating}
\begin{document}

\newcommand{\lsim}   {\mathrel{\mathop{\kern 0pt \rlap
  {\raise.2ex\hbox{$<$}}}
  \lower.9ex\hbox{\kern-.190em $\sim$}}}
\newcommand{\gsim}   {\mathrel{\mathop{\kern 0pt \rlap
  {\raise.2ex\hbox{$>$}}}
  \lower.9ex\hbox{\kern-.190em $\sim$}}}
\def\be{\begin{equation}}
\def\ee{\end{equation}}
\def\ba{\begin{eqnarray}}
\def\ea{\end{eqnarray}}
\def\d{{\rm d}}
\def\Ejet{E_{\rm jet}}
\def\tmin{t_{\rm min}}
\def\zmin{z_{\rm min}}
\def\Qmin{Q_{\rm min}}
\def\tsusy{t_{\rm SUSY}}
\def\Msusy{M_{\rm SUSY}}
\def\dlim{D_{\rm lim}}
\def\dmc{D_{\rm MC}}
\def\ap{\approx}

\title{\hfill {\small CERN-TH 2000-232}\\
       \vskip-0.4cm \hfill {\small IFIC/00-49}\\
       \vskip-0.4cm \hfill {\small hep-ph/0009053}\\
       \vskip1.0cm
       {\bf Monte Carlo simulation for jet fragmentation in SUSY-QCD}}

\author{V. Berezinsky$^{1,2}$ and M. Kachelrie{\ss}$^{3,4}$ \\ 
        {\it\small $^1$INFN, Laboratori Nazionali del Gran Sasso,
                   I--67010 Assergi (AQ), Italy} \\
        {\it\small $^2$Institute for Nuclear Research, Moscow, Russia}\\
        {\it\small $^3$TH Division, CERN, CH--1211 Geneva 23} \\
        {\it\small $^4$ Departament de F\'{\i}sica Te\`orica,
                   Universitat de Val\`encia - IFIC/CSIC, Spain} 
}

\date{September 1, 2000}
\maketitle

\abstract{
We present results from a new Monte Carlo simulation for jet fragmentation 
in QCD and SUSY QCD for large primary energies $\sqrt s$ up to $10^{16}$~GeV.
In the case of SUSY QCD the simulation takes into account not only gluons and
quarks as cascading particles, but also their supersymmetric partners.
A new model-independent hadronization scheme is developed, 
in which the 
hadronization functions are found from LEP data. An interesting
feature of SUSY QCD is the prediction of a sizeable flux of the lightest 
supersymmetric particles (LSPs), if R-parity is conserved. About $10\%$
of the jet energy is transferred to LSPs  which, owing to their harder spectra,
constitute an important part of the spectra for large $x=E/\Ejet$.
Spectra of protons and of secondary particles, photons and
neutrinos, are also calculated. These results have implications
for the decay of superheavy particles with masses up to the GUT scale, which
have been suggested as a source of ultrahigh energy cosmic rays. 
}


\def\baselinestretch{1.3}
\newpage

\section{Introduction}
QCD, being an essential part of the Standard Model, successfully
describes accelerator data for production of hadrons in $e^+e^-$
annihilation and in deep-inelastic scattering. There are two
distinctive parts in these calculations: the {\em perturbative QCD\/}
computation of the parton cascade in jets, and the parton {\em
hadronization}, in which low-virtuality partons are converted
non-perturbatively into hadrons.   

The QCD parton cascade is usually studied in Modified Leading Logarithmic 
Approximation (MLLA), where large logarithms, $\ln(Q^2)$ 
and $\ln(1/x)$, play a crucial role (here $Q^2$ is the maximum of
the perpendicular momentum $k_{\perp}$, and
$x=k_{\parallel}/k_{\parallel}^{\rm max}$).

This approximation is characterized by remarkable features. 

In the MLLA the QCD cascade has a probabilistic interpretation,
provided by the absence of interference terms in the tree diagrams. 
The colour coherence effect is taken into account in the MLLA.
It suppresses the emission of soft gluons and
results in the Gaussian peak of the parton distribution in terms of 
$\xi=\ln(1/x)$ (hump-backed plateau). 

The evolution of parton cascades in the MLLA (as well as in the LLA)
is adequately described by the
Dokshitzer--Gribov--Lipatov--Altarelli--Parisi (DGLAP) 
equations~\cite{DGLAP}. 

The parton spectra can be obtained {\em analytically\/} and by 
{\em Monte Carlo\/} (MC) simulations. 

Examples for analytical solutions are the {\em limiting 
spectrum\/}~\cite{lim}  
and the {\em Gaussian spectrum\/}~\cite{Gauss}, in which we include
the {\em distorted Gaussian spectrum\/}~\cite{dGauss,book}. The
limiting spectrum is the most accurate of them. Since we have a
special interest in it, we shall shortly review below the basic
assumptions under which this solution is obtained.

The limiting spectrum gives the energy spectrum of partons 
$D_{\rm lim}(\xi,Y)$ for a given center-of-mass energy $\sqrt{s}$ of
an $e^+e^-$-pair. Here $D$ is the number of cascade partons, 
$Y=\ln(\sqrt s/2\Lambda)$ and $\Lambda$ is the dimensional QCD scale. 

The analytical expression for $D_{\rm lim}(\xi, Y)$ is given in
Refs.~\cite{book,KhOc}. There are two fundamental parameters involved
in the limiting spectrum solution: the 
scale $\Lambda$ and the minimal virtuality $Q_0$ of
partons, down to which the cascade develops perturbatively; $Q_0$ can be
viewed as the effective mass of the partons. 
Two assumptions are necessary for the validity of the limiting
spectrum. The QCD coupling constant $\alpha_s(k_{\perp}^2)$ evolves
with $k_{\perp}^2$ effectively as in the one-loop 
approximation with three flavors $n_f=3$ {\em for all\/} $k_{\perp}^2$,
\be
\alpha_s(k_{\perp}^2)=\frac{12\pi}{(33-2n_f)\ln(k_{\perp}^2/\Lambda^2)} \,.
\label{alpha}
\ee
As a matter of fact, $\Lambda$ in Eq.~(\ref{alpha}) is treated in 
the limiting spectrum solution as a
free parameter to fit the $e^+e^-$-data. The best fit corresponds to 
$\Lambda =$ 250 -- 270~MeV. For this range of $\Lambda$ values, 
$\alpha_s(M_Z)$, given by Eq.~(\ref{alpha}) with $n_f=3$  is in the 
interval 0.118--0.120, to be compared with
the average experimental value $\alpha_s(M_Z)=0.1184 \pm 0.0031$
\cite{alpha}. Therefore, the 
phenomenological parameter $\Lambda$ coincides well with 
$\Lambda_{\rm QCD}$, which fits the experimental value of $\alpha_s(M_Z)$,
in the one-loop approximation with $n_f=3$.  

The second assumption, necessary for the derivation of the limiting
spectrum, is $Q_0=\Lambda$. It gives a reasonable value of $Q_0$, but 
the exact equality of these values has no theoretical justification.   

The limiting spectrum solution is valid only for small
$x\ll 1$. In this region, which includes the maximal values of 
multiplicity (in the Gaussian peak), it describes 
very accurately experimental data at all available energies $\sqrt s$ 
(see, e.g., \cite{KhLuOc,Lu98}). The large $x$ up to $x=1$ give the
dominant contribution to the total momentum of cascade partons. 
Therefore, the limiting spectrum solution does not guarantee that 
$\int xD_{\rm lim}(x,s)\d x$  precisely equals 2, although it can
be valid up to large $x$~\cite{book}.

{\em Monte Carlo\/} (MC) simulations of the QCD cascade give a more precise 
description of the cascade evolution. They are valid for all $x$ including  
$x \sim 1$. In contrast to the 
limiting spectrum, in MC simulations one can use 
$\alpha_s(k_{\perp}^2)$ with the measured value of
$\Lambda_{\rm QCD}$,  varying the number of flavours and  two-loop
corrections beeing taken into account. The assumption $\Lambda=Q_0$, specific
to the limiting spectrum, is not needed. MC simulations are based on a
probabilistic interpretation of the jet cascade.
Parton branching is described by the Altarelli--Parisi
functions, and the probability of parton evolution between two values of 
virtualities without branching is given by the Sudakov form factor. 
Finally, the coherent effect in the soft gluon emission 
(destructive interference) is conveniently taken into account by
angular ordering $\theta_1>\theta_2>\theta_3\ldots$ \cite{MaWe84}, 
where the indices number the 
generations (non-ordered processes are suppressed \cite{ord}). The first
MC simulation with angular ordering was performed in 
Ref.~\cite{MaWe84}. At present there are several detailed MC simulations, 
e.g. \cite{herwig,jetset,ariadne,isajet}, which differ mainly in
their description of hadronization.
We shall now briefly discuss the problem of hadronization. 

The description of parton hadronization is based on the assumption of 
Local Parton--Hadron Duality (LPHD) \cite{LPHD}. This hypothesis
implies that when $Q_0$ is small enough (of the order of $\Lambda$)
there is a proportionality between the spectra of partons and hadrons, with 
relations between their momenta, which are local in the phase space. Such
an interpretation can be justified by the idea of preconfinement
\cite{preconf}. 

As far as spectra are concerned, LPHD implies a proportionality between
the hadron and parton spectra. In Refs.~\cite{book,KhOc}, it is
emphasized that, most reasonably, this proportionality holds not on a
``one parton -- one hadron'' basis, but for the number of particles
averaged over a finite interval $\Delta \xi \sim 1$.

The LPHD hypothesis for limiting spectrum straightforwardly results 
\cite{book} in 
\be 
D_{\rm had}(x,\sqrt{s})=K_h(Q_0) D_{\rm part}(x,\sqrt{s},Q_0),
\label{K-h}
\ee
where the constant $K_h$ is universal, in the sense that it does not
depend on $\sqrt s$.

Equation~(\ref{K-h}) completes our description of the limiting spectrum, 
expressing the hadron spectrum through the spectrum of partons. 
The constant $K_h$, which connects the two spectra, 
is found from a comparison with experimental data as $K_h \approx 1.3$ 
for $\Lambda=Q_0\approx 270$~MeV \cite{KhLuOc}, and it does not 
change with energy unless some new physics (e.g. supersymmetry) appears. 

In MC simulations, the parameter $Q_0$ is in principle a free
parameter found by fitting experimental data. 
For HERWIG~\cite{herwig} and PYTHIA~\cite{jetset}, for example, $Q_0
\sim 1$~GeV.  
Several detailed hadronization models are used in simulations, e.g.
the independent fragmentation model, the Lund string model~\cite{lund}, 
and the cluster fragmentation model~\cite{cluster}. Usually, these
models use many free parameters and require to keep track 
of the four-momentum evolution of all partons.

The calculations described above are valid up to $\sqrt{s} \sim 1$--10~TeV. 
At higher energies the production of supersymmetric
particles is expected to change the results. One
might be interested  
in much higher energies, being inspired by the production of superheavy 
particles up to the GUT scale in the Universe. Such particles can be produced
by topological defects and by many processes at the post-inflationary 
stage of the Universe. Recently superheavy particles with masses 
$M_X \sim 10^{12}$--$10^{14}$~GeV attracted much attention as a source
of the observed Ultra High Energy Cosmic Rays (UHECR) with 
energies $10^{19}$--$10^{20}$~eV (for recent reviews, see \cite{SHP}).

The limiting spectrum for SUSY QCD was calculated in Ref.~\cite{BeKa98} 
for very high energies $\sqrt{s}$, corresponding to masses of superheavy
particles $M_X \sim 10^{12}$--$10^{14}$~GeV. The supersymmetric
partons (squarks and gluinos or jointly spartons) participate in the
cascade until the virtualities $t$ of the particles drop below the
mass scale of SUSY particles, $t\sim M_{\rm SUSY}^2$. Then a SUSY
particle decays, producing in the end the Lightest Supersymmetric
Particle (LSP), for which the lightest neutralino is usually  considered. 
The role of supersymmetric partners is
two-fold: they double the number of parton types in the cascade, and they
change the evolution of $\alpha_s(k_{\perp}^2)$. Even at small $t \ll
M_{\rm SUSY}^2$, the cascade remembers the number of flavours at large $t$
because, for example, each
squark leaves after its decay a quark, which continues QCD cascading. At
large $t$ and small $x \ll 1$ gluons and gluinos dominate and their 
``children'' constitute the dominant part of the cascade at small
$t$. Therefore, the dominant contribution to the limiting spectrum is
given by gluons and gluinos.  

The SUSY QCD limiting spectrum solution has two drawbacks with respect
to ordinary QCD. First, the number of flavours that determine the
evolution of the coupling constant according to Eq.~(\ref{alpha})  
has to be fixed to {\em one\/} value of $n_f$ for the whole
range of $k_\perp^2$. Second, the limiting spectrum for ordinary QCD
is normalized by experimental data, which are absent in the case of SUSY QCD.
Normalization due to the conservation of momentum 
$\int x D_{\rm lim}(x,s)\d x=2$ is unreliable, since the
limiting spectrum is not valid for large $x$, which give the main
contribution to the integral (see discussion in \cite{BeKa98}).

During the last few years, the production and decays of supersymmetric
particles have been included in most MC simulations focusing on LHC studies. 
Although the LHC will operate above the expected threshold of
SUSY particle production, its
energy is not large enough for these particles to 
participate in the QCD cascade. Therefore, all currently available MC 
simulations consider only
on-shell decays of spartons  and neglect possible 
branchings of
gluinos and squarks\footnote{The future C++ version of HERWIG will
  include branchings of spartons~\cite{Webber}.}. 
Another obstacle against the use of standard MC simulations at
extremely large energies around $\sqrt s \sim 10^{12}-10^{14}$~GeV is 
that the necessary numerical precision and the required amount of
memory space and computing time become a challenge for present--day 
computers.

We have therefore developed a new MC simulation, which includes as
cascading particles not only gluons and quarks but also gluinos and
squarks. We consider cascades that are initiated by the decay to two
jets of superheavy particles with mass $M_X \sim 10^{12}$--$10^{14}$~GeV, 
or by $e^+e^-$ annihilation at $s=M_X^2$. SUSY partons, 
squarks and gluinos, are produced in the fragmentation of ordinary partons
and vice versa. All squarks and gluinos are assumed to have equal
masses, for which we use $M_{\rm SUSY}=200$~GeV and $M_{\rm SUSY}=1000$~GeV.
When the virtuality of the cascading particles drops below $M_{\rm SUSY}^2$, 
spartons decay to LSPs (neutralinos), which freely escape. 
The perturbative development of the cascade continues with
ordinary partons until their virtualities 
reach $Q_0^2$, for which we use $Q_0^2=0.625$~GeV$^2$ to fit
the data at 
small energies $\sqrt{s}$. We use a new hadronization procedure. It is based 
on a model-independent, phenomenological approach, in which hadronization
functions for large $s$ (or $M_X$) are calculated from hadron spectra 
observed at small $s$ ($M_X$). This method can be used for any type
of hadrons as well as for photons and neutrinos, if their spectra 
are known with good enough accuracy at small energy. 
 
Following Ref.~\cite{book1}, we shall use the following notation:\\
$\Lambda$ is the dimensional QCD scale,\\
$Y=\ln(\sqrt{s}/2\Lambda)$,\\
$t=p_{\mu}^2$ is the virtuality of cascade partons,\\
$Q^2=t_{\rm max}$ is the virtuality of the primary parton,\\
$Q_0^2$ is the minimum virtuality of the perturbative evolution
of the QCD cascade,\\ 
$z=E'/E$, where $E$ and $E'$ are the energies of ingoing and outgoing
partons at fragmentation,\\
$\zeta=1-\cos\theta$, where $\theta$ is the angle between two outgoing 
partons,\\
$\tilde{t}=\zeta E^2$,\\
$k_{\perp}, k_{\parallel}$ are the transverse and parallel momenta
transferred, respectively,\\
$x= k_{\parallel}/k_{\parallel}^{\rm max}$, \\
$\xi=\ln(1/x)$.

\section{MC Simulation of the Perturbative Phase of SUSY QCD Cascades}
The perturbative part of our simulation is very similar to those of
MCs for ordinary QCD cascades, except for including spartons and the
condition for their exit from the cascade.
We consider a superheavy $X$ particle with mass $M_X$ which decays
into two jets with energy $\Ejet=M_X/2$. We assume that the primary
partons produced in the $X$ particle decay have the maximum virtuality
$Q^2=m_X^2/4$ and that the $X$ particle has equal branching ratios to
all partons. As to the first assumption, in reality, there is 
a distribution of partons with different $t$, but the Sudakov form factors 
suppress small $t$ values. The second assumption is made because of  
the unspecified interactions of the $X$ particles.

Our simulation closely follows the angular ordered parton cascade
algorithm  developed in Refs.~\cite{MaWe84,cluster}. It is 
convenient to use in this algorithm the variable 
$\tilde t=\zeta E^2$,  where $E$ is the energy of the incoming parton, 
$\zeta\ap 1-\cos\theta$, and $\theta$ is the angle between the two
emitted partons. A primary parton with energy $\Ejet$ ($=m_X/2$)
and angular variable $\xi_0\leq 1$ initiates a cascade, which proceeds 
until the ordinary partons reach the minimal virtuality $\tilde t = 4Q_0^2$. 
Here the perturbative evolution of the cascade terminates.

In each branching of an incoming parton $i$ with $\tilde t'$, we generate with
the veto algorithm~\cite{jetset-manual} a new $\tilde t$  and $z$
according to the probability distribution  
\be  \label{P}
 \d{\cal P}_i(\tilde t,z)= 
 \sum_{jk} \frac{\d \tilde{t}}{\tilde t} \frac{\d z}{2\pi} \, 
 \alpha_{s}\left[ z^2 (1-z)^2 \tilde{t}  \right] P_{i\to jk}(z) \,
 \frac{\Delta_i(\tilde{t}')}{\Delta_i(\tilde{t})}  \,.
\ee
Here, the sum includes all possible branching channels $jk$, 
$z^2 (1-z)^2 \tilde{t}$ is the parton transverse momentum, and
$\Delta_i$ is the product of the individual Sudakov-like form factors 
 $\Delta_{i\to jk}$ \cite{cluster} , 
\be     \label{sudakov}
\Delta_{i\to jk} (\tilde t) = \exp \left[ 
  - \int_{4\tilde{t}_{\rm min}}^{\tilde t}  \frac{\d t'}{t'} \,
f_{i\to jk} 
(t') \right] 
\ee
with
\be     \label{sudakov2}
 f_{i\to jk} (\tilde t) = 
    \int_{z_{\rm min}}^{z_{\rm max}} \frac{\d z}{2\pi} \, 
    \alpha_{s}\left[ z^2 (1-z)^2 \tilde t \right]  P_{i\to jk}(z) \,.
\ee
The unregularized Altarelli--Parisi splitting functions $P_{i\to  jk}(z)$ 
of SUSY QCD \cite{splitt} are given in
Table~\ref{splitting_functions}. 

The angular ordering $\zeta_j, \zeta_k <\zeta_i$ for the branching $i \to jk$,
which takes into account colour coherence, is equivalent to 
$\tilde{t}_j< z^2\tilde{t}_i$ and $\tilde{t}_k < (1-z)^2\tilde{t}_i$. 
These conditions result in 
\be
 z_{\rm min} = \sqrt{\tilde{t}_{\rm min}/\tilde t} \,, \qquad
 z_{\rm max} = 1- \sqrt{\tilde{t}_{\rm min}/\tilde t} \,.
\ee

For the evolution of the running coupling $\alpha_s$ as a function of
gluon virtuality $t$ at small momentum transfer $t < t_{\rm SUSY}$, we use 
the standard two-loop dependence with variable $n_f$ and thresholds, 
and normalize  $\alpha_s$ as $\alpha_s(M_Z)=0.119$, which corresponds to   
$\Lambda_{\overline{MS}}^{(5)}=222$~MeV. At large momentum transfer 
$t>t_{\rm SUSY}$ we use minimal SUSY-SU(5) coupling constant evolution 
\cite{GUT}, normalizing the coupling constant at 
$\sqrt t=M_{\rm GUT}= 1\cdot 10^{16}$~GeV, as 
$\alpha_s(M_{\rm GUT}^2)\ap1/25.8$. Explicitly we use
\be   \label{alpha>}
 \alpha_s(t) = \frac{\alpha(M_{\rm GUT}^2)}
                    {1+ b_s/(4\pi) \ln(t/M_{\rm GUT}^2)\alpha(M_{\rm GUT}^2)},
\ee
where $b_s= 9-n_f$ is a constant, that governs the evolution of the 
coupling constant with $t$. At $t>t_{\rm SUSY}$, $n_f=6$ and $b_s=3$.
The above assumption means that we introduce, instead of many
thresholds corresponding to SUSY particles with different masses, a single
threshold at $t=t_{\rm SUSY}$. This is a reasonable thing to do in
view of the large uncertainties in our knowledge of mass spectrum of
SUSY particles. Equation~(\ref{alpha>})  
approximates accurately enough the evolution of $\alpha_s(t)$ as
calculated in Ref.~\cite{la93}, when 
$t_{\rm SUSY} \approx 2\cdot 10^5~$GeV$^2$.
Starting from this value, $\alpha_s(t)$ evolves in the regime of 
Eq.~(\ref{alpha>}). Note that $t_{\rm SUSY}$ does not necessarily coincide 
with the scale
$M_{\rm SUSY}$, the universal mass of squarks and gluino, for which we use
as two representative values $M_{\rm SUSY}= 200$~GeV and $M_{\rm SUSY}=1$~TeV.
In particular, the low value of $t_{\rm SUSY}$ 
used here is compatible with much larger $M_{\rm SUSY}$, as emphasized
in Ref.~\cite{la93}.

Finally, we have to specify the value of the cut-off $\tilde\tmin$ for the
cascade evolution. We do not distinguish between different quark
flavours and use $\tilde\tmin=0.625$~GeV$^2$ for all branchings in which
only ordinary particles are produced and $\tilde\tmin=M_{\rm SUSY}^2$, where
$M_{\rm SUSY}$ is the typical mass scale of the spartons, for
branchings in which SUSY particles are produced, respectively.

Let us now describe a step $i\to jk$ in our simulation. 
For an  incoming parton $i$ with $\tilde{t}'$ we generate first a new cascade 
variable $\tilde t$, according to the probability distribution given
by the ratio
$\Delta_i(\tilde{t}')/\Delta_i(\tilde t)$. Then we select the
branching channel $jk$ using as weight $f_{i\to jk}(\tilde t)$ and generate 
$z$ according to the probability distribution 
$\alpha_{s}\left[ z^2 (1-z)^2 \tilde t \right]  P_{i\to jk}(z)$.

The last ingredient in the perturbative part of our simulation is the
exit of supersymmetric particles from the cascade. We assume that the
neutralino $\tilde{\chi}$ is the LSP and that R-parity is conserved. 
Reaching $\tilde{t}_{\rm min}=M_{\rm SUSY}^2$, squarks and gluinos decay as 
$\tilde q \to q+\tilde{\chi}$ and $\tilde g \to q+\bar q+\tilde{\chi}$,
thus producing UHE LSPs. 

In fact, we are running in this work two Monte Carlo similations: with
the ordinary QCD and with SUSY QCD. 

In the former case supersymmetric partons are not included, and for
perturbative calculations we assume the SM particle content with 
$\alpha_s(t)$ evolution in two-loop approximation with proper
thresholds. We fix $Q_0^2=0.625$~GeV$^2$. We need these calculations
mostly to test our method. 

The assumptions of SUSY QCD Monte Carlo are described above. At 
$\tilde{t}_{min}<M_{SUSY}^2$ cascade develops according to ordinary
QCD scheme.

\section{Hadronization}

The Monte Carlo simulation described in the last section is completely 
determined by perturbative physics. How the spectrum of coloured
quarks and gluons $D_i(x,\sqrt{s})$ is transformed into the  spectrum
of hadrons
$D_{\rm had}(x,\sqrt{s})$ is still an open problem. Monte Carlo
simulations have to use some hadronization model (see Introduction),
which describes the non-perturbative evolution of the cascade for
$\tilde t<4\tilde t_{\rm min}$. Two hadronization models, the cluster
fragmentation model \cite{cluster} used in HERWIG~\cite{herwig} and
the Lund string model~\cite{lund} used in PYTHIA~\cite{jetset}, 
require the knowledge of the
four-momenta of all the partons. Thus these models need detailed time
and memory-consuming computations.

We suggest here a phenomenological, model-independent hadronization
scheme based on the knowledge of the hadron spectra at energies
$\sqrt{s}$ smaller than the energy of interest.
This method is valid for any hadron type and can be applied 
to the secondary particles, such as photons and neutrinos, as well.
The application of this method is somewhat restricted (e.g. it cannot
give the angular distribution of particles in a jet or correlations),
but its use is very efficient for the decay of superheavy particles, 
where multiplicity, and hence
the number of partons to follow in a simulation is very large.

Our hadronization scheme depends on only one theoretical
assumption, which is reliable and testable.  Namely, we assume that 
the unknown non-perturbative physics
can be factorized into hadronisation functions $f_i(z)$ that
do {\em not\/} depend on $\sqrt{s}$, 
\be  \label{hadron}
 D_h (x,\sqrt{s}) = \sum_{i=q,g} \int_x^1 \frac{\d z}{z} \:  
                          D_i (x/z, \sqrt{s}) f_i^h (z) \,,
\ee
where the index $h$ runs through different types of hadrons,
e.g. $\pi^0, \pi^\pm, N$ etc.

The functions $f_i^h(z)$ give the probability that a parton $i$  with
energy $E$ is converted into a hadron $h$ with energy $zE$. It is
implicitly assumed in Eq.~(\ref{hadron}) that the perturbative cut-off
$Q_0$ is fixed, and $f_i^h$ is determined for this value of $Q_0$,
although in principle for every $Q_0$ and $D_i(x,\sqrt{s},Q_0)$ one can
find $f_i^h(z,\sqrt{s},Q_0)$ to fit the observed hadron spectra.

Equation~(\ref{hadron}) with energy independent hadronization functions
follows from basic principles and is confirmed (see below) at energies
of $e^+ e^-$ colliders.
It has the form of a Volterra integral equation of
the first kind though in contrast to the standard case, the RHS contains
not one but two unknown functions $f_i$ for every $h$. In principle,
the two functions 
$f_g(x)$ and $f_q(x)$ can be uniquely determined if $D_h$ is known 
as an analytic function without errors for two different values of
$\sqrt{s}$. In practice, $D_h(x)$ is known only as a discrete set
of experimental data and Eq.~(\ref{hadron}) represents an ill-posed
inversion problem%
\footnote{Volterra integral equations of the first kind can be solved  
  normally by linearization, even if the LHS are data. However, the lower 
  integration limit in (\ref{hadron}) does not represent a sharp cut-off
  because the kernels $D_i(x)$ vanish for $x\to 1$. Therefore, 
  Eq.~(\ref{hadron}) behaves effectively like a Fredholm equation, and
  these are known to be extremely ill-conditioned.}~\cite{cr86}.
Instead of solving Eq.~(\ref{hadron}) by an inversion method, we prefer to
find physically motivated trial functions for $f_i$ to fit the experimental
data at $\sqrt{s}=91.2$~GeV. 

In terms of the more convenient variable $\xi=\ln(1/x)$,
Eq.~(\ref{hadron}) has the form 
\be  \label{hadron_l}
 D_h (\xi, \sqrt{s}) 
 = \sum_{i=q,g} \int_0^l \d\xi' \:  D_i (\xi-\xi', \sqrt{s}) f_i (\xi') \,,
\ee
where the index $h$ in the hadronization functions is suppressed.

In the limiting spectrum, when $Q_0=\Lambda$, the hadronization functions
$f_i$ are proportional to delta functions. Inspired by this analytical
solution we choose for $f_i$  Gaussian functions, 
\be
 f_{i}(\xi)=a_i \exp\left(- \frac{(\xi-\xi_{{\rm max},i})^2}{\sigma_i^2}
                  \right) \,.
\ee 
With this hadronization function the approximate proportionality holds
between spectra of partons and hadrons as LPHD demands. The position of
the peak in the hadronization function determines the shift between the
maxima of parton and hadron spectra. While for gluons the
hadronization function $f_g(\xi)$ should vanish for $\xi\to 0$,
because gluons have to split their energy to a $q\bar q$ pair, for quarks
$f_q(\xi)$ can be finite at $\xi=0$. 

\begin{figure}
\begin{picture}(15,9)
 \put(2.5,8.3) { \epsfig{file=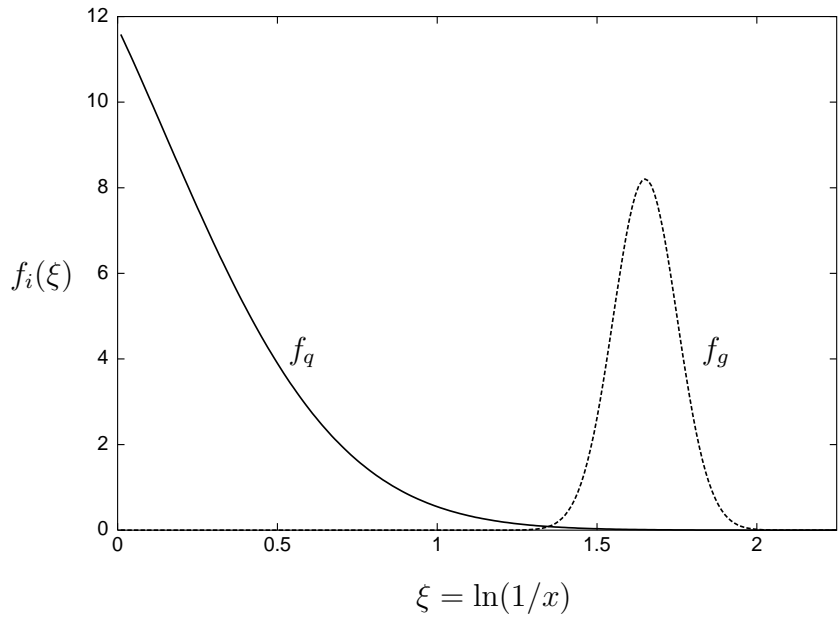,height=10.5cm,width=7.5cm,angle=270} }
 \put(7.2,0.2) {$\xi=\ln(1/x)$}
 \put(1.8,4.5) {$f_i(\xi)$}
 \put(5.5,3.5) {$f_q$}
 \put(11.,3.5) {$f_g$}
\end{picture}
\caption{\label{f_had} 
Hadronization functions for quarks $f_q(\xi)$ (solid line) and 
for gluons $f_g(\xi)$ (broken line) 
obtained by fitting Gaussians to experimental data at $\sqrt{s}=91.2$~GeV.} 
\end{figure}

The hadronization functions we obtained for $Q_0^2=0.625$~GeV$^2$ from
a fit to LEP data at $\sqrt{s}=91.2$~GeV are shown in 
Fig.~\ref{f_had}.

Our hadronization scheme has been tested by two methods: for
relatively small energies, $\sqrt{s}=58$~GeV and 133~GeV, we
confronted our calculations with LEP data, and for very large
$\sqrt{s}$ (or $M_X$) we compared the calculated spectrum with the
limiting spectrum, using a special case 
when it is correct (see below). In both cases ordinary QCD Monte Carlo
was used. 

Figures 2-4 display a comparison between the charged hadron spectrum
from our MC simulation for ordinary QCD and experimental data \cite{data} at 
$\sqrt{s}=58$, 91.2 and 133~GeV, respectively.

Let us now discuss whether the hadronization functions $f_i(\xi)$ found
from the fit to data at $\sqrt{s}=91.2$~GeV can really be used at
$M_X=10^{12}$--$10^{16}$~GeV. 

First of all we note, that a test can be given by LPHD, which demands 
approximate
proportionality between parton and hadron spectra. It implies that the
$\xi'$ values, that give the dominant contribution to the integral in
Eq.~(\ref{hadron_l}), are about the same at $\sqrt{s}=91.2$~GeV and at
large $M_X$. Numerical tests show that this is indeed the
case for both the quark and the gluon contribution.

As a critical test of our hadronization scheme, we compared 
the limiting spectrum with the results of our simulation for a special,
well-controlled case of ordinary QCD with the number of quark flavours
$n_f=3$, and with $\alpha_s(k_{\perp}^2)$ given by Eq.(\ref{alpha}).
For the limiting spectrum in this case we can use the normalization
constant $K_{h}\ap 1.3$ obtained by
fitting experimental data \cite{Lu98}.  Since we do not introduce
new high-energy physics, the limiting spectrum is valid for any
initial energy $M_X$ (see Introduction). 
In Fig.~\ref{r_13}, we show the ratio of these two (charged) hadron spectra.
The agreement between the two spectra
is excellent, except for the small $\xi\lsim 6$ region where it is
known that the limiting spectrum is not valid.  The disagreement
reaches $50\%$ at $\xi \approx 2.1$ ($x \approx 0.12$).

In conclusion, we think that our hadronization recipe is a valid
alternative to the extrapolation of the Lund string or the cluster 
fragmentation model to extremely large $M_X$.

\section{Results: Spectra of Hadrons and Secondary Particles}
Using the algorithm for the perturbative evolution of the
SUSY QCD cascade as 
described in Section~2 and our hadronization scheme from Section~3, 
we can now compute the fragmentation spectra of hadrons. 
As numerical values for $M_X$, we choose in the 
graphs given as examples three values interesting for UHECR physics, 
$M_X=10^{12},10^{13},10^{14}$~GeV, 
as well as $M_X=10^6$ and $10^{16}$~GeV as lowest and highest scale of
interest.  Similarly, we use $M_{\rm SUSY}=200$~GeV and 
$M_{\rm SUSY}=1000$~GeV as two representative values for the SUSY mass
scale.

In Figs.~\ref{had_200} and \ref{had_1000}, the hadron
spectra $\d N_{\rm had}/\d\xi$ from SUSY QCD MC simulations 
are displayed as function of $\xi$ for 
$M_{\rm SUSY}=200$~GeV and $M_{\rm SUSY}=1000$~GeV, respectively; in
both figures the spectra were calculated for $M_X=10^{12}, 10^{13},
10^{14}$~GeV. For the GUT scale $M_X=10^{16}$~GeV, the hadron spectra
$\d N_{\rm had}/\d\xi$ are shown in Fig.~\ref{had_M16} and for the low
scale $M_X=10^6$~GeV in Fig.~\ref{had_M6}. 
{\em The hadron spectra depend only weakly on $M_{\rm SUSY}$,}
with increasing differences for larger values of $M_X$.
Both effects are easy to understand: when spartons disappear from 
the cascade at $\tilde t\sim\Msusy^2$ due to on-shell decays, each of
them leaves there an ordinary parton with similar virtuality.
Therefore, the cascade proceeds as if nothing had happened, except
that some energy is lost through the emission of neutralinos and
leptons, which is not large ($\sim 10\%$).  
Second, the importance of spartons for the cascade decreases for
smaller values of $M_X$, thereby reducing also the dependence of the
hadron spectra on $\Msusy$ for smaller $M_X$.

The weak dependence of the spectra on $M_{SUSY}$ justifies our
choice of an universal value for the masses of supersymmetric particles. 
Indeed, if we assume now that supersymmetric particles have different
masses in the range 200-1000~GeV, the resulting hadron spectra will
differ less than in Figs.~\ref{had_M16} and \ref{had_M6}.

The signature of supersymmetry in decays of superheavy $X$ particles
is the production of LSPs,  which we assume as stable
neutralinos. They are generated in the cascade mostly when the
virtuality of the spartons approaches $M_{\rm SUSY}^2$. The calculated
neutralino spectra are shown in Figs.~\ref{LSP_200}--\ref{LSP_M6}
for the same parameters as the hadron spectra in 
Figs.~\ref{had_200}--\ref{had_M6}. 
Like the hadron spectra, they have  the characteristic
Gaussian form, however with a shifted position of their maxima due to 
their larger cut-off $\Msusy$ in the shower development. The energy
fraction taken away by the neutralinos is typically $10\%$ for values
of $M_X$ interesting for UHECR physics, with a minimum of $5\%$ for 
$M_X=10^{6}$~GeV and $\Msusy=1$~TeV and a maximum of  $12\%$ for 
$M_X=10^{16}$~GeV and $\Msusy=200$~GeV.

We have only derived a common hadronization function for all hadrons and, 
consequently, we cannot calculate directly, e.g. pion or nucleon spectra
through Eq.~(\ref{hadron}). 
Since the fraction of energy $\epsilon_i$ going into
different meson and baryon species is determined by the
non-perturbative process of hadronization, these fractions as the
hadronization functions themselves do not depend on $s$. Thus, 
we can use the value from $Z$ decay, $\epsilon_N\ap 0.05$ and
$\epsilon_\pi\ap 0.95$. Then
\be  \label{nucl}
 \frac{\d N_{\rm nucl}}{\d x} = 
               \epsilon_N   \frac{\d N_{\rm had}}{\d x}\,,\qquad\quad
 \frac{\d N_\pi}{\d x}        = \epsilon_\pi \frac{\d N_{\rm had}}{\d x} \,.
\ee

Using the hadron spectra obtained in the last Section,
it is simple to calculate analytically the spectra of
secondary particles, photons and neutrinos. 
The normalized photon spectrum from a decay of one $X$ particle at rest 
is given by
\be \label{photon}
 \frac{\d N_\gamma}{\d x} = \frac{2}{3} \epsilon_\pi \int_x^1  
                            \frac{\d y}{y}\:\frac{\d N_{\rm had}}{\d y} \,.
\ee

The total neutrino spectrum, given by the sum from decays of pions and
muons, can be presented in the following form,
\be \label{neutr}
 \frac{\d N_{\nu}}{\d x} = \frac{2}{3} \epsilon_\pi \left(
                \frac{\d N_{\nu_\mu}}{\d x} (\pi\to\mu\nu_\mu) + 
                \frac{\d N_{\nu_\mu}}{\d x} (\mu\to\nu_\mu\nu_e e) + 
                \frac{\d N_{\nu_e}}{\d x} (\mu\to\nu_\mu\nu_e e) \right)\,,
\ee
where for pion decay
\be
 \frac{dN_{\nu_\mu}}{\d x}(\pi \to \mu\nu_{\mu}) = 
 R \int_{Rx}^1  \frac{\d y}{y}\: \frac{\d N_{\rm had}}{\d y}
\ee
and for muon decay
\be  \label{muon}
 \frac{\d N_{\nu_i}}{\d x}(\mu \to \nu_{\mu}\nu_e e) = 
  R \int_x^1 \frac{\d y}{y} \int_y^{y/r} \frac{\d y'}{y'}\:
             \frac{\d N_{\nu_i}}{\d y} \, \frac{\d N_{\rm had}}{\d y'} \,,
\ee
with
\be
 \frac{\d N_{\nu_e}}{\d y}=2-6y^2+4y^3 \,, \qquad\quad
 \frac{\d N_{\nu_{\mu}}}{\d y}=\frac{5}{3}-3y^2+\frac{4}{3}y^3,
 \label{nu-e}
\ee
and $r=(m_{\mu}/m_{\pi})^2$, $R=1/(1-r)$.

The resulting nucleon, photon and neutrino spectra $x^3 \d N_i/\d x$ are
shown as functions of $x$ together with the spectra of neutralinos in
Fig.~\ref{frag} for $\Msusy=200$~GeV and $M_X=10^{12}$~GeV and 
$M_X=10^{14}$~GeV, respectively. We have multiplied the
spectra by $x^3$  in order to facilitate the comparison of our spectra
with the energy spectra of observed UHECR. At $x\gsim 0.7$,
the spectra have some uncertainties because of the unknown branching
ratios of the $X$ particle into (s)partons and fluctuations due to the
small number of produced particles. The excess of nucleons over
secondary particles, neutrinos and gamma, at largest $x$ is a result
of kinematical effect and the steep spectra of pions and
nucleons in the end of the spectrum.

\section{Discussion}

In this Section we compare for large $M_X$ the results of our MC for
the two cases, SUSY QCD and ordinary-QCD, with
other computations, and most notably with limiting spectrum calculations.  
The latter case of ordinary QCD is formally a special case of our MC
simulation for SUSY QCD in the limit $\Msusy,~\tsusy\to\infty$,
i.e. $\alpha_s$ is given by two-loop approximation with variable $n_f$, 
and the probability to produce a sparton is zero.

The validity of our method has been proved by the tests described in 
Section 3. If no new physics beyond the three light quark flavours is
introduced, the $k_{\perp}$-dependence of $\alpha_s$ is given by
Eq.~(\ref{alpha})  with $n_f=3$ and the limiting spectrum with
$K_h=1.3$ is valid for arbitrary high energies. We can calculate the
hadron spectrum in our ordinary QCD MC (hadronization procedure
included), introducing there the same assumptions about $n_f$ and
$\alpha_s$. The excellent agreement is illustrated by Fig.~5. The
disagreement seen at large $x$ is natural, because the limiting
spectrum is not valid there. 

It is instructive to compare our MC for ordinary QCD with variable $n_f$
and the exact behaviour of $\alpha_s(k_{\perp})$ with the limiting spectrum 
with fixed number of flavours $n_f=3$ and $n_f=6$. It is clear that, 
in neither case,
$\alpha_s(k_{\perp})$ from Eq.~(\ref{alpha}) describes correctly
$\alpha_s$ in the whole interval of $k_{\perp}$, and the MC spectrum
should be between these two solutions. Figures.~6 and 7 show that
this is indeed the case. The accuracy of each limiting spectrum 
compared with the MC spectra is better than 30--50\%. 

In Ref.~\cite{sarkar}, HERWIG was used to obtain fragmentation spectra
in case of ordinary QCD. The maximal mass $M_X$ possible to simulate 
was $M_X=10^{11}$~GeV and even for this not very large value of 
$M_X$ the computations required several months. The spectra were displayed 
only for large $x>0.01$, beyond the Gaussian peak. One of the conclusions 
of this work was that at $x \geq 0.2$ the proton yield is higher than
the photon and neutrino yield. However, it was later realized that
this result is caused by the tendency of HERWIG to overproduce protons at
large $x$ (Ref.~\cite{Rubin}, see also \cite{sarkar2}).

Let us come over to our SUSY QCD MC and compare the simulated spectra
with the SUSY limiting spectrum \cite{BeKa98}. The spectra disagree 
both in the position of the Gaussian peak and in its height. To clarify
which assumptions of the SUSY QCD limiting spectrum are responsible
for this disagreement, we re-run the SUSY QCD MC simulation with a set of
assumptions  as similar as possible to those used in the
derivation of the SUSY QCD limiting spectrum. We found that 
the main reason for the disagreement is the universal dependence 
of $\alpha_s(t)$, taken as $\alpha_s^{-1}(t)=(b_s/4\pi)\ln(t/\Lambda^2)$,
with $b_s=3$ for SUSY, together with $\Lambda=Q_0=250$~MeV.
It differs from $\alpha_s$ with a variable number of flavours, which 
is used in SUSY QCD MC, by a
factor 1.4--3 in the whole $k_\perp^2$ range, with largest disagreement at
small $k_{\perp}$. Changing the evolution of $\alpha_s(k_{\perp})$, 
an agreement can be reached between the MC and the SUSY QCD limiting spectrum:
we run the SUSY QCD MC including only gluons and (massless) gluinos
with fixed $b_s=3$ and with frozen $\alpha_s(\tilde t)$ for $\tilde
t<0.9$~GeV$^2$, which is a reasonable physical assumption.
The comparison with
the SUSY QCD limiting spectrum for partons is shown in
Fig.~\ref{gg_p_12}. The two spectra agree indeed quite well.

An interesting alternative approach to computing the fragmentation
spectra produced  
by decays of superheavy particles was suggested in a recent
work~\cite{Rubin}. In this method, the event generator
SPYTHIA~\cite{spythia} was used to 
simulate fragmentation spectra of partons  and spartons into
protons, photons, and neutrinos at the scale $M_X=10^4$~GeV. 
Then the DGLAP equations were used to evolve the
fragmentation functions up to the scale $10^{12}$--$10^{13}$~GeV.

It is premature to compare our results, since in \cite{Rubin}
preliminary results are presented, but the spectra, as displayed 
in \cite{Rubin} and \cite{sarkar2}, do not agree well with ours. 
In particular, the Gaussian peak is broader than in our calculations, 
Fig.~\ref{frag}. Comparing these spectra  
one should be aware of the differences in methods and assumptions. 
For example, we treat spartons as cascading particles, while in SPYTHIA 
spartons are taken as on-shell particles, which decay but do not cascade.
On the other hand, the SPYTHIA spectrum is used only as the input, 
and the evolution to higher energies includes cascading. This
difference will be eliminated with the C++ version of
HERWIG~\cite{Webber}, which will be available soon.

\section{Summary}
We have developed a new MC simulation for jet fragmentation in ordinary QCD 
and SUSY QCD, which is valid for initial energies up to the GUT scale.
The simulation includes a perturbative part, operating at
virtualities higher than the infrared cut-off $Q_0^2=0.625$~GeV, and 
a hadronization part. 

The perturbative part for SUSY QCD includes squarks and gluinos as
cascade particles with a universal mass $M_{\rm SUSY}$. The evolution 
of $\alpha_s(t)$ takes into account the correct number of active
flavours and spartons at a given $t$. 
The influence of the scale $M_{\rm SUSY}$ on the hadron spectrum is rather
weak for the studied range $300 \leq M_{\rm SUSY}\leq 1000$~GeV. It implies
that if supersymmetric particles have different masses in the range 
200 -1000~GeV, the resulting difference in the hadron spectra remains small 
(see Section IV and Figs.~\ref{had_M16} and \ref{had_M6}).

The hadronization scheme is model-independent and based on the well
justified and tested assumption that the hadronization function
$f_i^h(z)$, see Eq.~(\ref{hadron}), does not depend on $\sqrt s$. 
Thus the hadronization function could be calculated from LEP data. Our
scheme was tested at $\sqrt{s}=58$ and 133 GeV against
experimental data and
for very large values of $M_X$ by a comparison with the
limiting spectrum solution for ordinary QCD. 
For the aim of this comparison, we calculated hadron spectra using the
MC for ordinary QCD in the case when the limiting spectrum is known to
be correct: $n_f=3$ and $\alpha_s(t)$ given by Eq.~(\ref{alpha}).
The excellent agreement between both spectra is illustrated by 
Fig.~\ref{r_13}.

The spectra of nucleons and secondary particles, photons,
neutrinos, as well as neutralinos, have been calculated and presented 
in Fig.~\ref{frag}. These spectra can be used for calculations of
fluxes of ultra high energy  cosmic rays, produced by superheavy dark
matter and by topological defects.

\section*{Acknowledgments}
We acknowledge with gratitude the participation of R. Sang in the 
initial stage of this work. 
We are grateful to Yu.L. Dokshitzer, V.A. Khoze, S.S. Ostapchenko and
B.R. Webber for many useful explanations and comments.
MK would like to thank the Alexander von Humboldt-Stiftung for 
a Feodor-Lynen grant and the EC for a Marie-Curie grant.
The work of MK was partially also supported by DGICYT grant PB95-1077 and
by the EC under the TMR contract ERBFMRX-CT96-0090. The work by VB was 
performed within INTAS Project No 99-01065.

\newpage

\begin{table}
\begin{center}
\begin{tabular}{c|c}
splitting channel $i\to jk$ & splitting function $P_{i\to jk}(z)$\\
\hline 
$$ & $$  \\[-1.0ex]
$g \to g + g$&
        $3 \,\left[{z \over 1-z} + {1-z \over z} + z (1-z) \right]$ \\[1.5ex]
$g \to \tilde g + \tilde g$&
        $3 \,\left[ z^2 + (1-z)^2 \right]$ \\[1.5ex]
$g \to q + q$&
        $\frac{n_f^\ast}{2} \,\left[ z^2 + (1-z)^2 \right]$ \\[1.5ex]
$g \to \tilde q + \tilde q$&
        $3 \,\left\{ 1 - [z^2 + (1 - z)^2] \right\}$ \\[1.5ex]
$\tilde g \to g + \tilde g$&
        $3 \,{1 + (1-z)^2 \over z}$ \\[1.5ex]
$\tilde g \to \tilde q + q$&
        $3 \, z$ \\[1.5ex]
$q \to q + g$ &
        ${4\over 3} \,{1 + z^2 \over 1-z} $ \\[1.5ex]
$q \to \tilde q + \tilde g$&
        ${4\over 3} \,{z} $ \\[1.5ex]
$\tilde q \to \tilde q + g$&
        ${4\over 3} \, \left[{1 + z^2 \over 1 -z} - (1-z) \right] $ \\[1.5ex]
$\tilde q \to q + \tilde g$&
        ${4\over 3}$ \\
\end{tabular}
\end{center}
\caption{Splitting functions $P_{i\to jk}(z)$, where $z$ is the energy
fraction of the particle $j$.
\label{splitting_functions}}
\end{table}

\phantom{}
\newpage
\phantom{}
\unitlength1.0cm

\begin{figure}
\begin{picture}(15,9)
 \put(2.5,8.3) { \epsfig{file=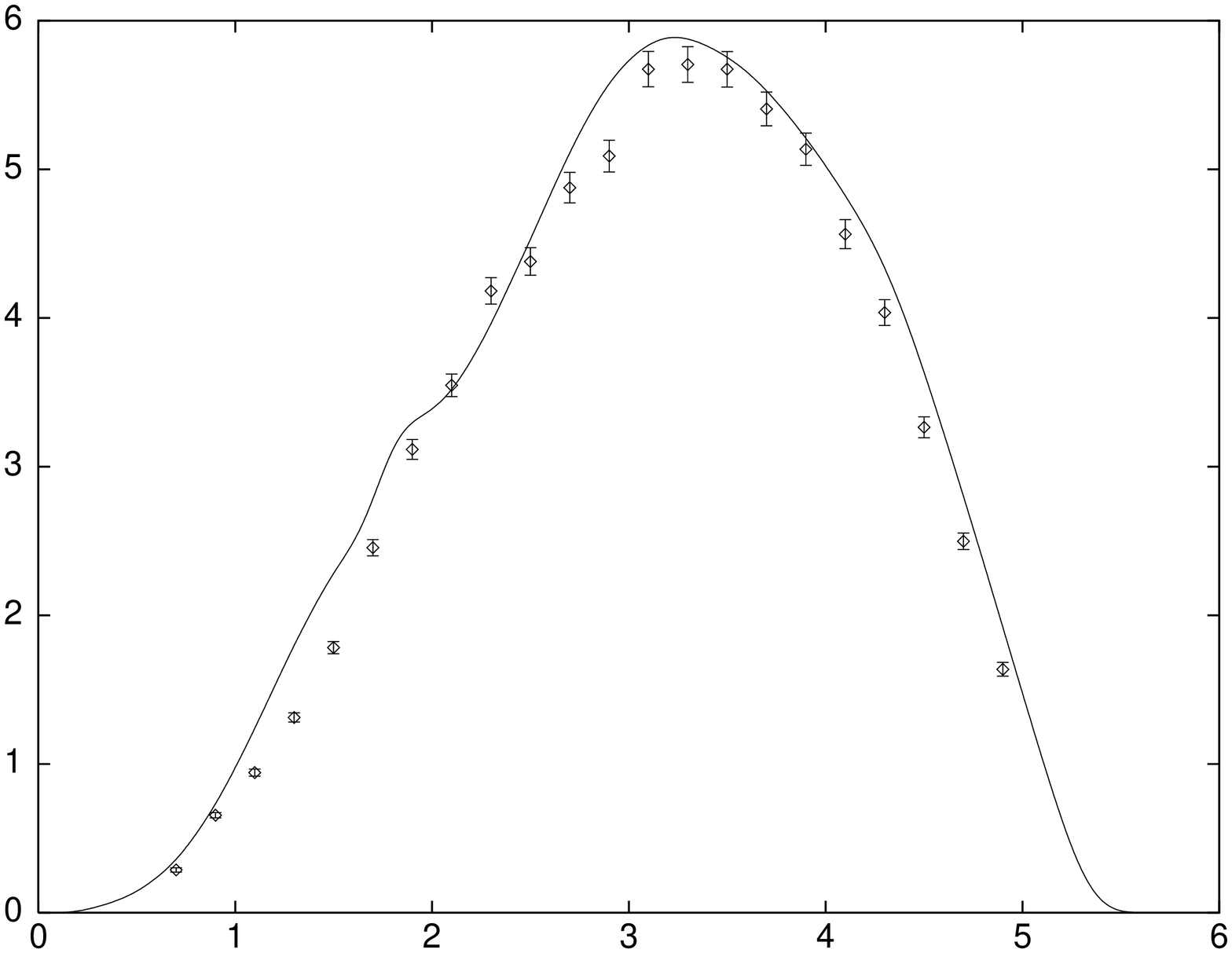,height=10.5cm,width=7.5cm,angle=270} }
 \put(7.2,0.2) {$\xi=\ln(1/x)$}
 \put(1.8,4.5) {{\Large $\frac{\d N_{\rm ch}}{\d\xi}$}}
\end{picture}
\caption{\label{had_58}
Comparison of the spectrum of charged hadrons 
$\d N_{\rm ch}/\d\xi$ from ordinary QCD Monte Carlo simulation (solid line)
with the experimental data (shown with errorbars) at $\sqrt{s}=58$~GeV.} 
\end{figure}

\begin{figure}
\begin{picture}(15,9)
 \put(2.5,8.3) { \epsfig{file=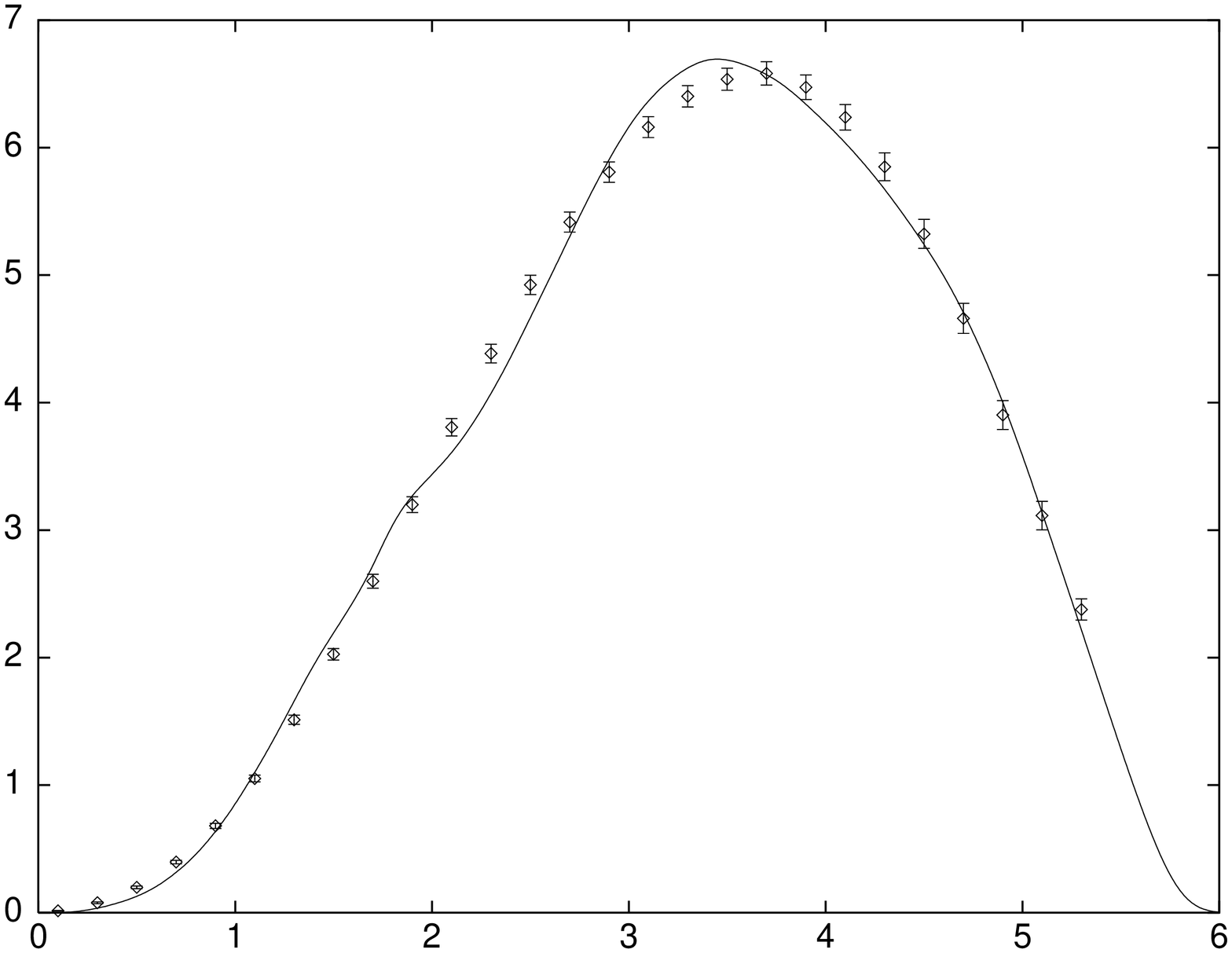,height=10.5cm,width=7.5cm,angle=270} }
 \put(7.2,0.2) {$\xi=\ln(1/x)$}
 \put(1.8,4.5) {{\Large $\frac{\d N_{\rm ch}}{\d\xi}$}}
\end{picture}
\caption{\label{had_91}
Comparison of the spectrum of charged hadrons 
$\d N_{\rm ch}/\d\xi$ from ordinary QCD Monte Carlo simulation (solid line)
with the experimental data (shown with errorbars) at $\sqrt{s}=91.2$~GeV.} 
\end{figure}

\begin{figure}
\begin{picture}(15,9)
 \put(2.5,8.3) { \epsfig{file=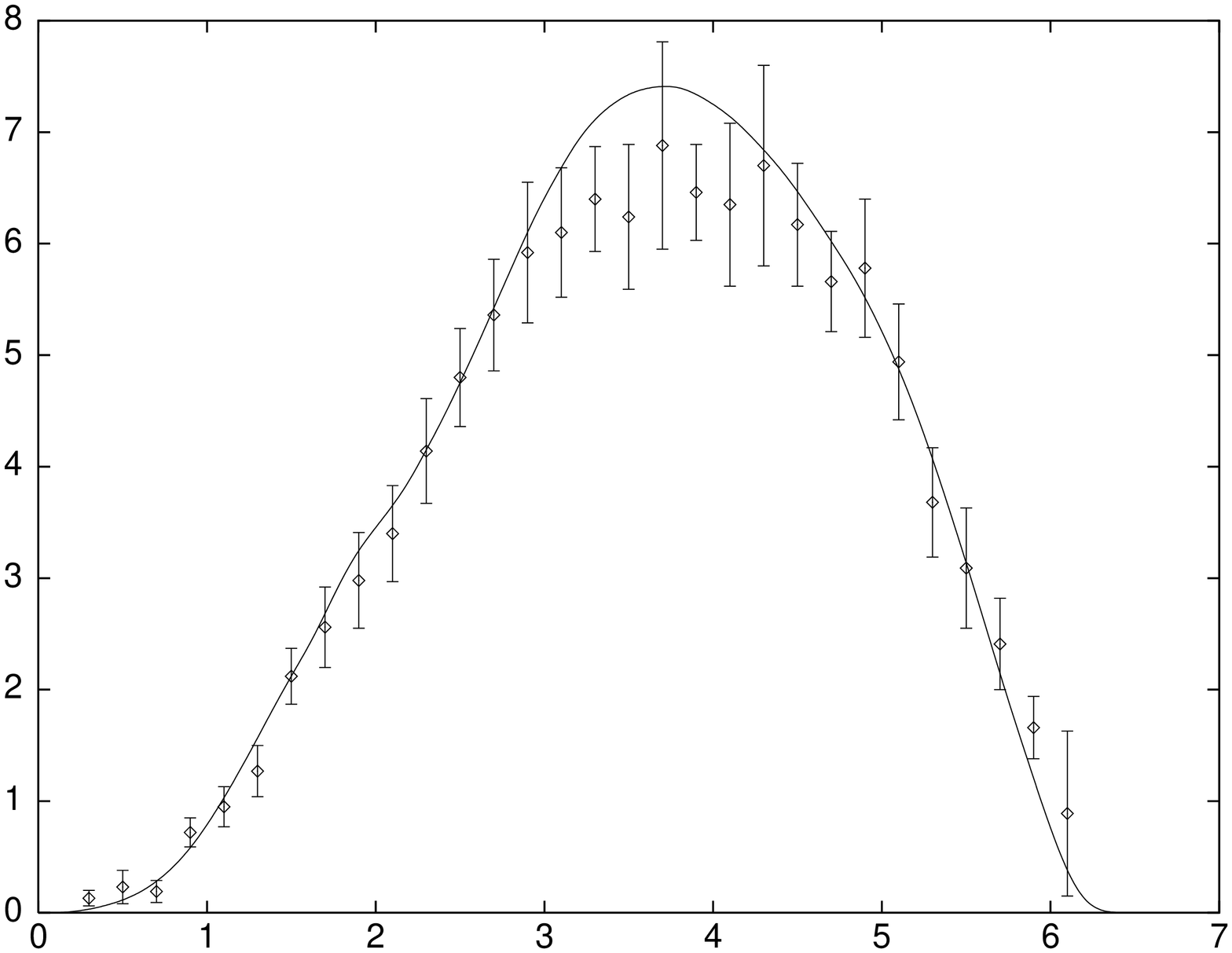,height=10.5cm,width=7.5cm,angle=270} }
 \put(7.2,0.2) {$\xi=\ln(1/x)$}
 \put(1.8,4.5) {{\Large $\frac{\d N_{\rm ch}}{\d\xi}$}}
\end{picture}
\caption{\label{had_133} 
Comparison of the spectrum of charged hadrons 
$\d N_{\rm ch}/\d\xi$ from ordinary QCD Monte Carlo simulation (solid line)
with the experimental data (shown with errorbars) at $\sqrt{s}=133$~GeV.} 
\end{figure}


\begin{figure}
\begin{picture}(15,9)
 \put(2.5,8.3) { \epsfig{file=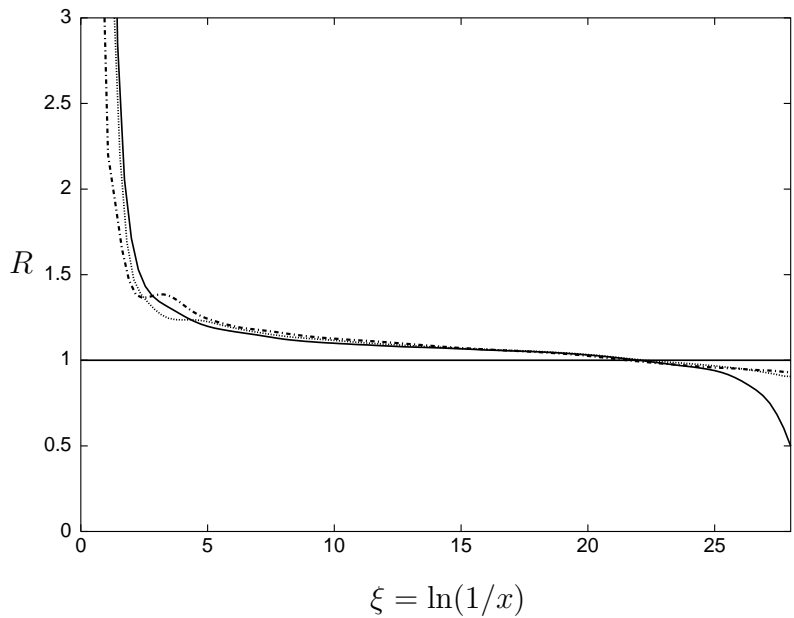,height=10.5cm,width=7.5cm,angle=270} }
 \put(7.2,0.2) {$\xi=\ln(1/x)$}
 \put(2.4,4.65) {$R$}
\end{picture}
\caption{\label{r_13}
The ratio $R=D_{\rm lim}(\xi)/D_{\rm MC}(\xi)$ 
of the limiting spectrum and of the hadron spectrum from
the simulation for $M_X=10^{12}$~GeV (solid line), $10^{13}$~GeV
(broken line) and $10^{14}$~GeV (dashed line). 
All for ordinary QCD with $n_f=3$.}
\end{figure}


\begin{figure}
\begin{picture}(15,9)
 \put(2.5,8.3) { \epsfig{file=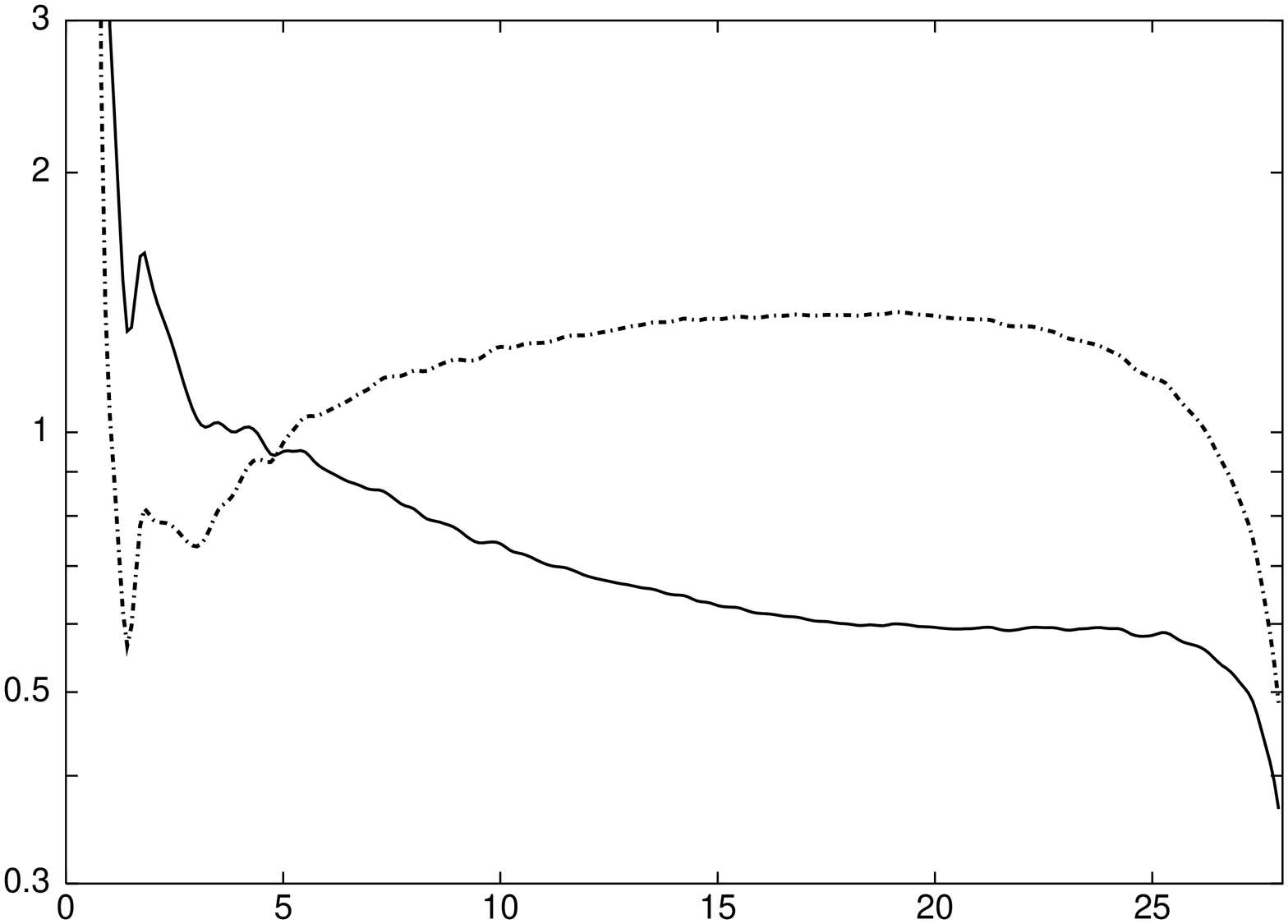,height=10.5cm,width=7.5cm,angle=270} }
 \put(7.2,0.2) {$\xi=\ln(1/x)$}
 \put(2.4,4.65) {$R$}
\end{picture}
\caption{\label{r_n_12}
Comparison of the limiting spectrum of QCD for $n_f=3$ and $n_f=6$ 
with the ordinary QCD spectrum from the Monte Carlo simulation:
$R=\dlim({\rm QCD},n_f=3)/\dmc$ (solid line) and
$R=\dlim({\rm QCD},n_f=6)/\dmc$ (broken line). Both for $M_X=10^{12}$~GeV.}
\end{figure}

\begin{figure}
\begin{picture}(15,9)
 \put(2.5,8.3) { \epsfig{file=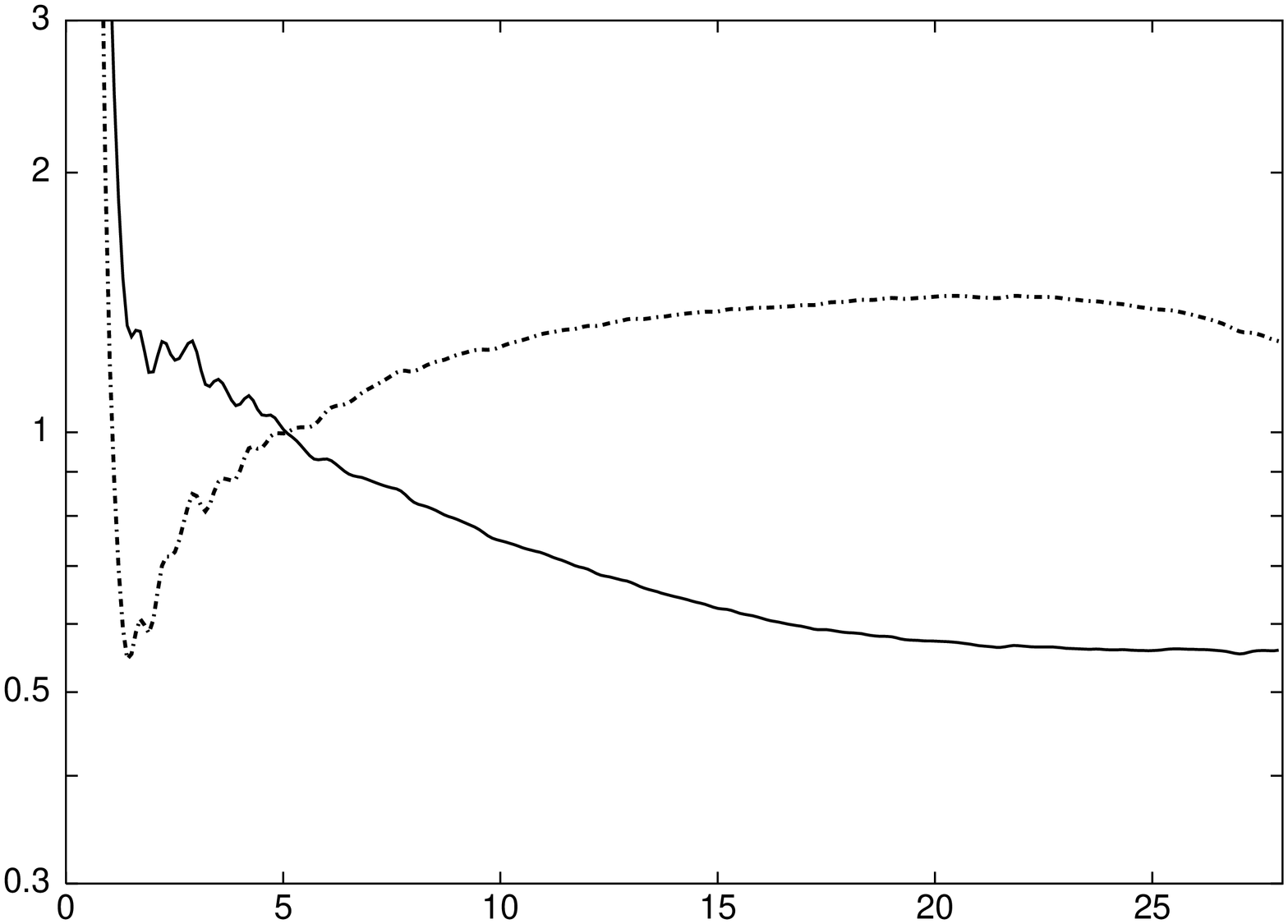,height=10.5cm,width=7.5cm,angle=270} }
 \put(7.2,0.2) {$\xi=\ln(1/x)$}
 \put(2.4,4.65) {$R$}
\end{picture}
\caption{\label{r_n_14}
Comparison of the limiting spectrum of QCD for $n_f=3$ and $n_f=6$ 
with the ordinary QCD spectrum from the Monte Carlo simulation:
$R=\dlim({\rm QCD},n_f=3)/\dmc$ (solid line) and
$R=\dlim({\rm QCD},n_f=6)/\dmc$ (broken line). Both for $M_X=10^{14}$~GeV.}
\end{figure}


\begin{figure}
\begin{picture}(15,9)
 \put(2.5,8.3) { 
               \epsfig{file=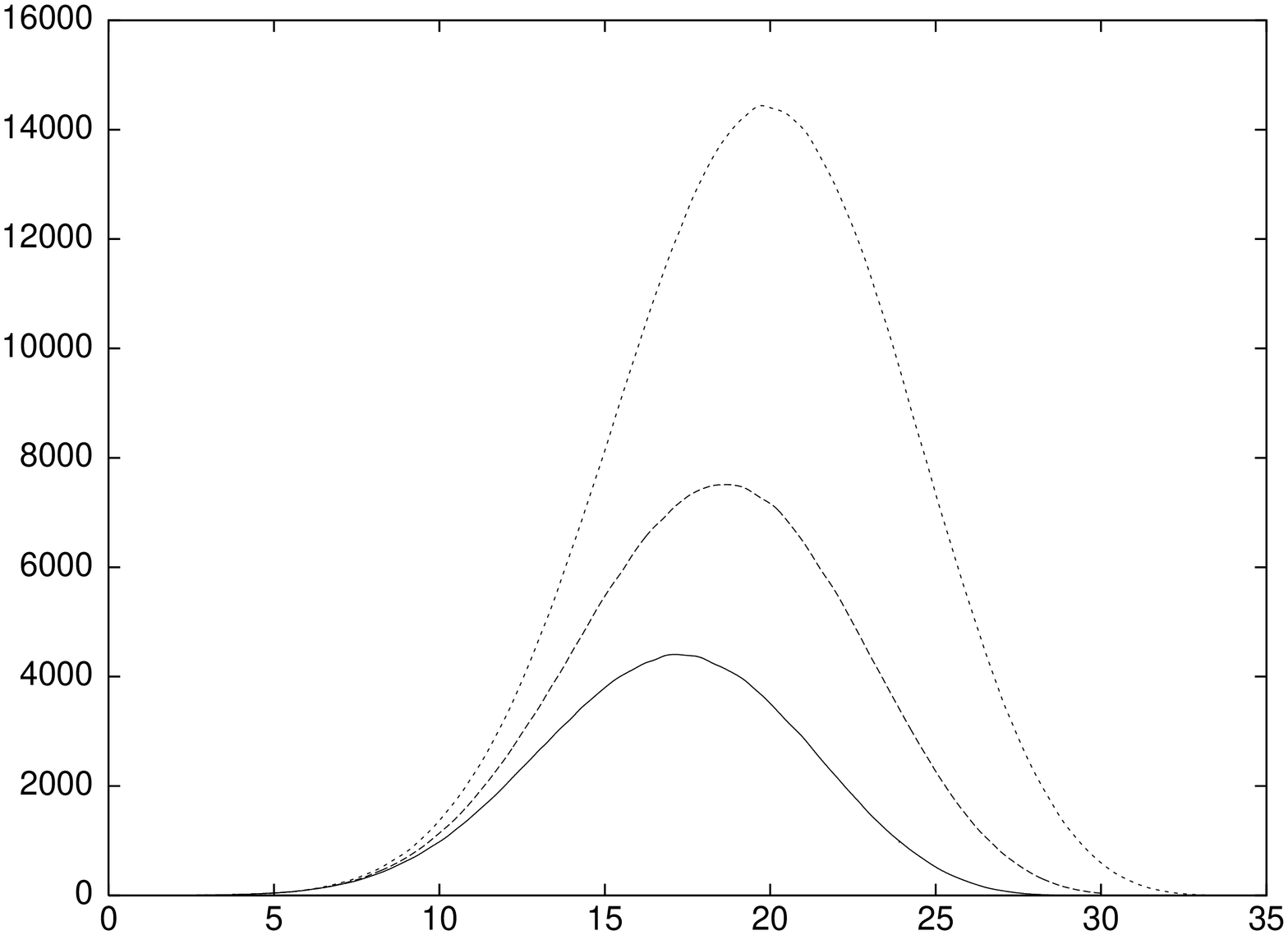,height=10.5cm,width=7.5cm,angle=270} }
 \put(7.2,0.2) {$\xi=\ln(1/x)$}
 \put(1.5,4.5) {{\Large $\frac{\d N_h}{\d\xi}$}}
\end{picture}
\caption{\label{had_200}
Hadron spectra $\d N_h/\d\xi$ from SUSY QCD Monte Carlo simulation 
for $M_X=10^{12}$~GeV (bottom),
$M_X=10^{13}$~GeV (middle) and $M_X=10^{14}$~GeV (top), 
all for $\Msusy=200$~GeV.}
\end{figure}

\begin{figure}
\begin{picture}(15,9)
 \put(2.5,8.3) { 
               \epsfig{file=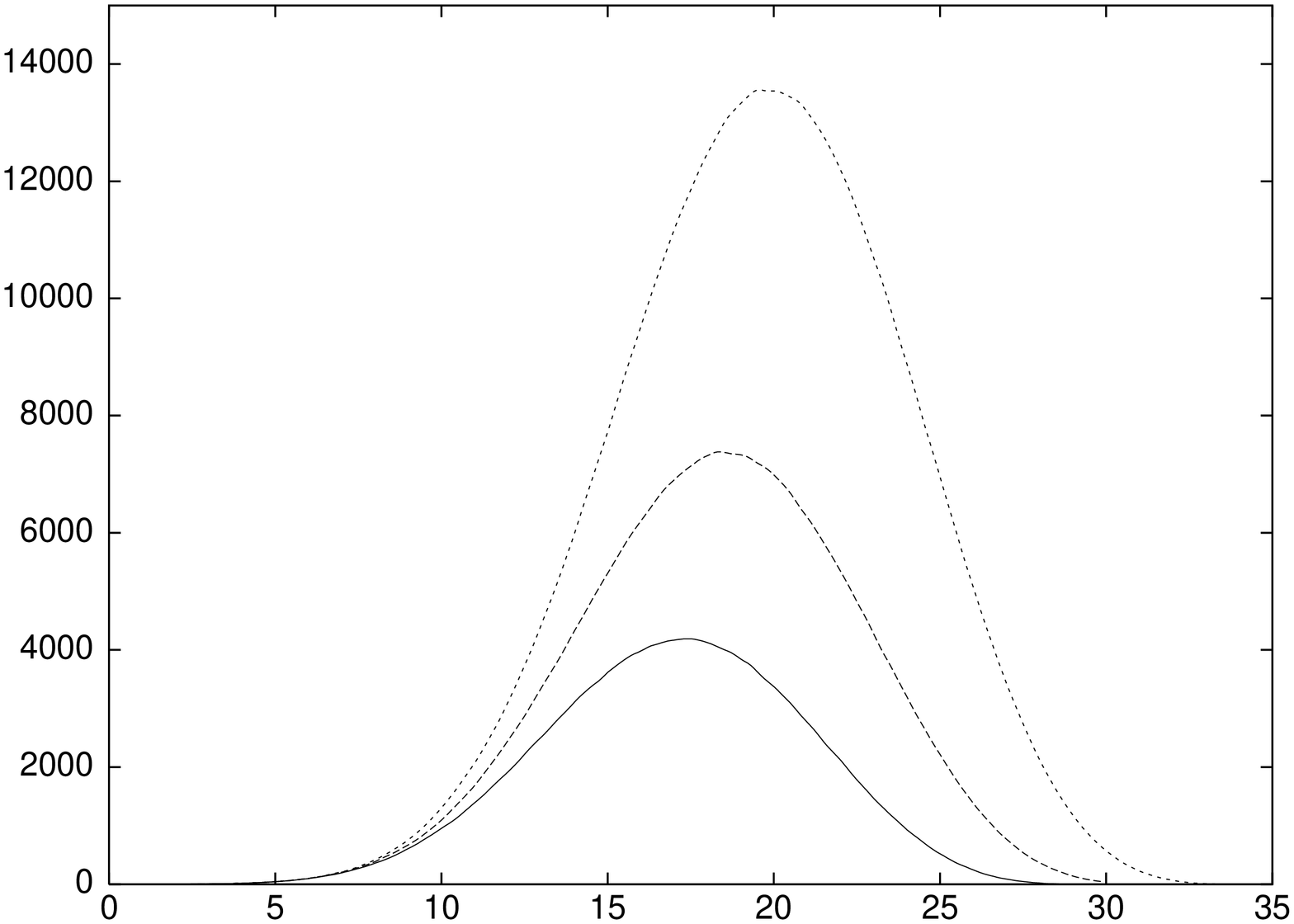,height=10.5cm,width=7.5cm,angle=270} }
 \put(7.2,0.2) {$\xi=\ln(1/x)$}
 \put(1.5,4.5) {{\Large $\frac{\d N_h}{\d\xi}$}}
\end{picture}
\caption{\label{had_1000}
Hadron spectra $\d N_h/\d\xi$ from SUSY QCD Monte Carlo simulation
for $M_X=10^{12}$~GeV (bottom),
$M_X=10^{13}$~GeV (middle) and $M_X=10^{14}$~GeV (top), 
all for $\Msusy=1$~TeV.} 
\end{figure}

\begin{figure}
\begin{picture}(15,9)
 \put(2.5,8.3) { 
               \epsfig{file=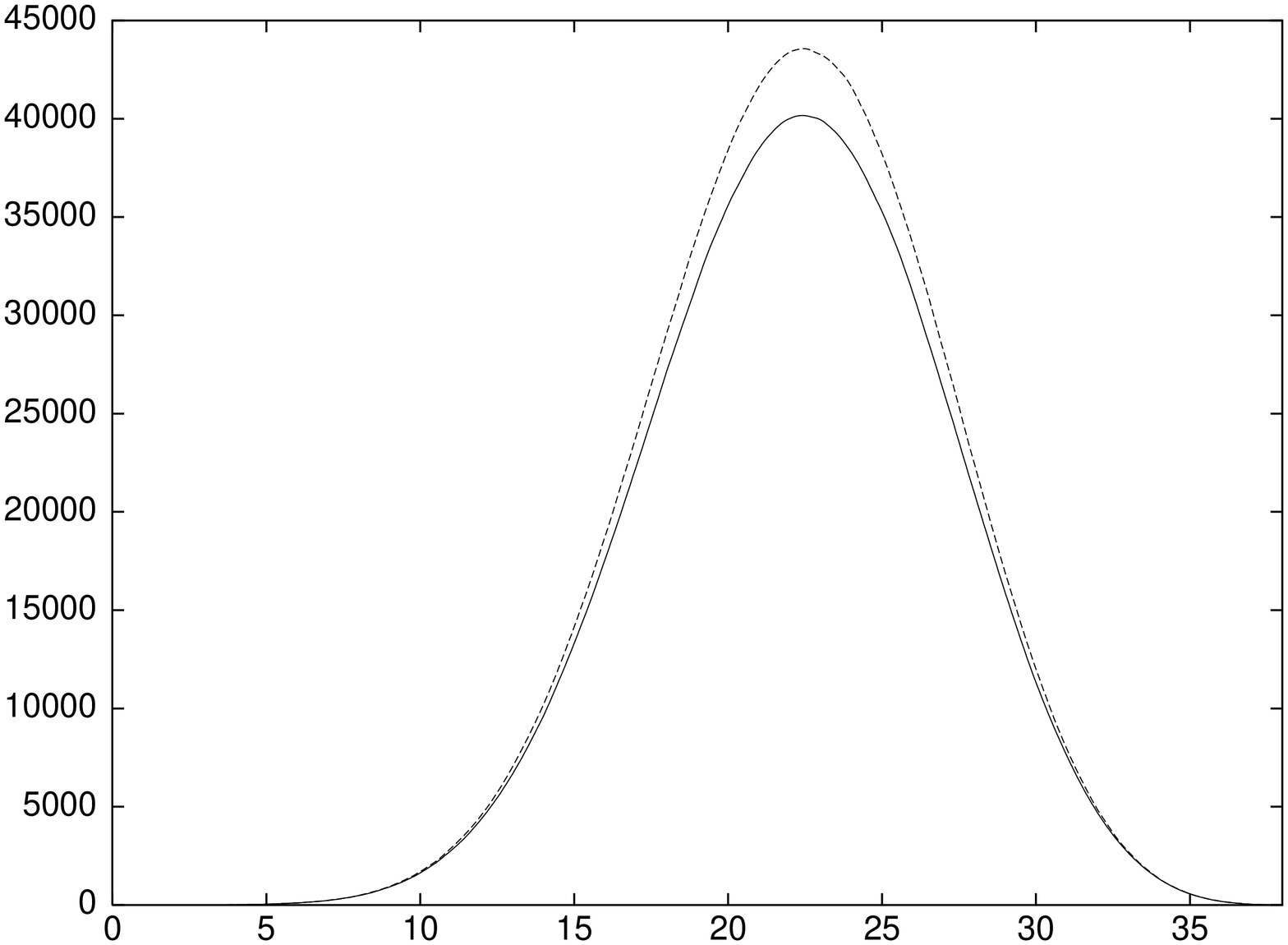,height=10.5cm,width=7.5cm,angle=270} }
 \put(7.2,0.2) {$\xi=\ln(1/x)$}
 \put(1.5,4.5) {{\Large $\frac{\d N_h}{\d\xi}$}}
\end{picture}
\caption{\label{had_M16}
Hadron spectra $\d N_h/\d\xi$ for SUSY QCD Monte Carlo simulation
for $\Msusy=200$~GeV (broken line) 
and $\Msusy=1$~TeV (solid line), both for $M_X=10^{16}$~GeV.}
\end{figure}

\begin{figure}
\begin{picture}(15,9)
 \put(2.5,8.3) { 
               \epsfig{file=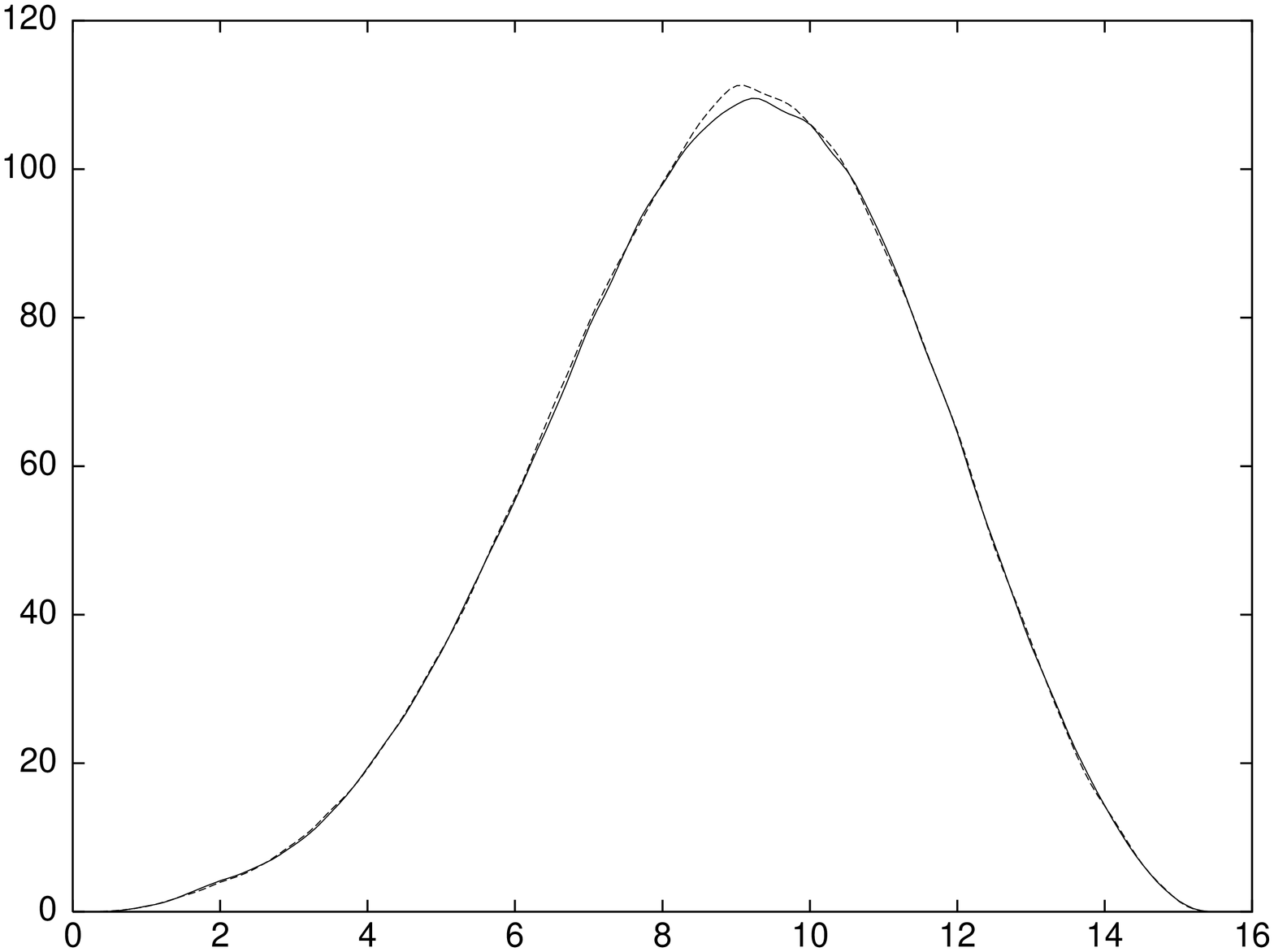,height=10.5cm,width=7.5cm,angle=270} }
 \put(7.2,0.2) {$\xi=\ln(1/x)$}
 \put(1.5,4.5) {{\Large $\frac{\d N_h}{\d\xi}$}}
\end{picture}
\caption{\label{had_M6}
Hadron spectra $\d N_h/\d\xi$ from SUSY QCD Monte Carlo simulation 
for $\Msusy=200$~GeV (broken line) and 
$\Msusy=1$~TeV (solid line), both for $M_X=10^{6}$~GeV.}
\end{figure}


\begin{figure}
\begin{picture}(15,9)
 \put(2.5,8.3) { 
         \epsfig{file=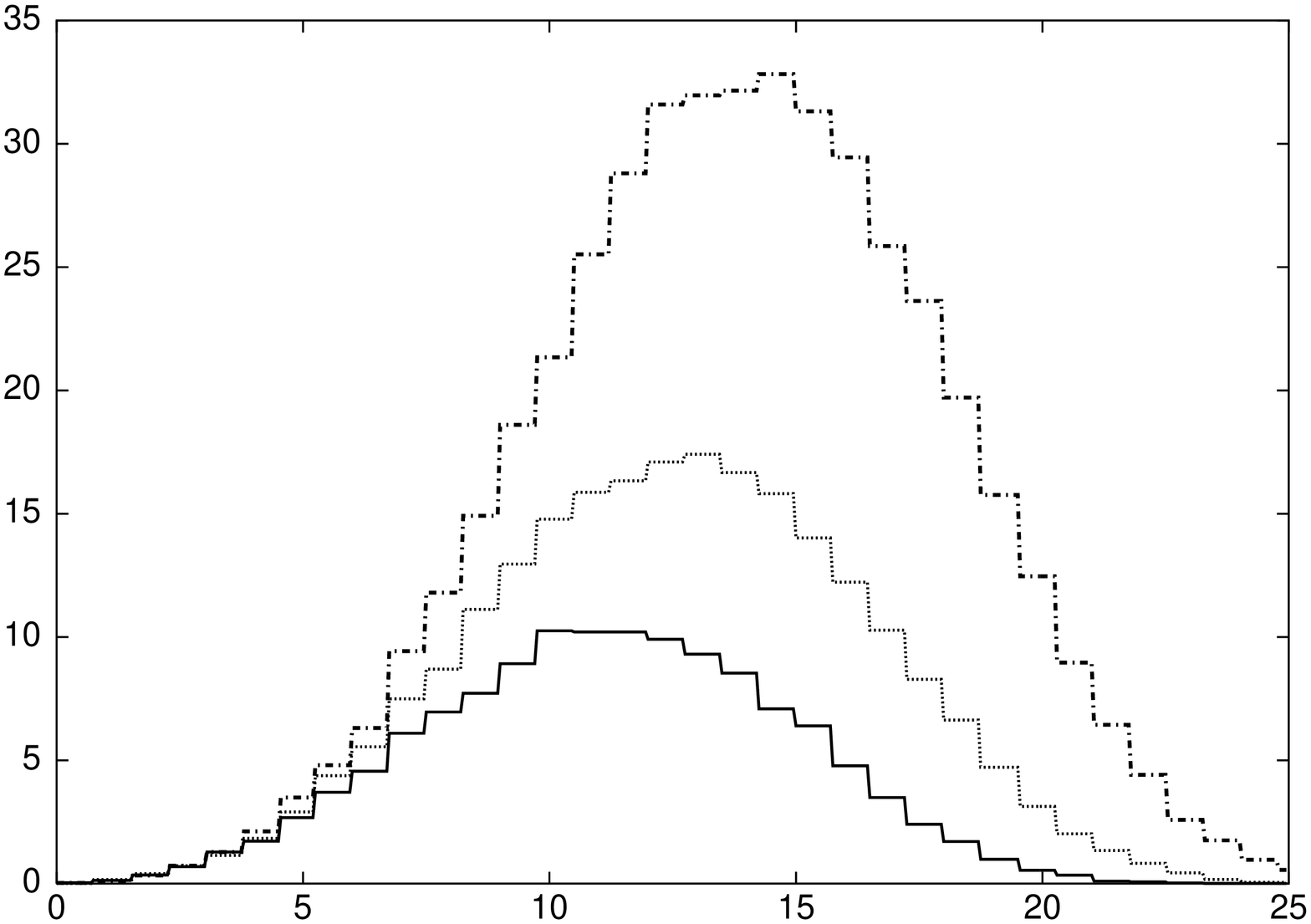,height=10.5cm,width=7.5cm,angle=270} }
 \put(7.2,0.2) {$\xi=\ln(1/x)$}
 \put(1.8,4.5) {{\Large $\frac{\d N_\chi}{\d\xi}$}}
\end{picture}
\caption{\label{LSP_200}
Neutralino spectra from SUSY QCD Monte Carlo simulation for 
$M_X=10^{12}$~GeV (bottom),
$M_X=10^{13}$~GeV (middle) and
$M_X=10^{14}$~GeV (top), 
all for $\Msusy=200$~GeV.}
\end{figure}

\begin{figure}
\begin{picture}(15,9)
 \put(2.5,8.3) { 
         \epsfig{file=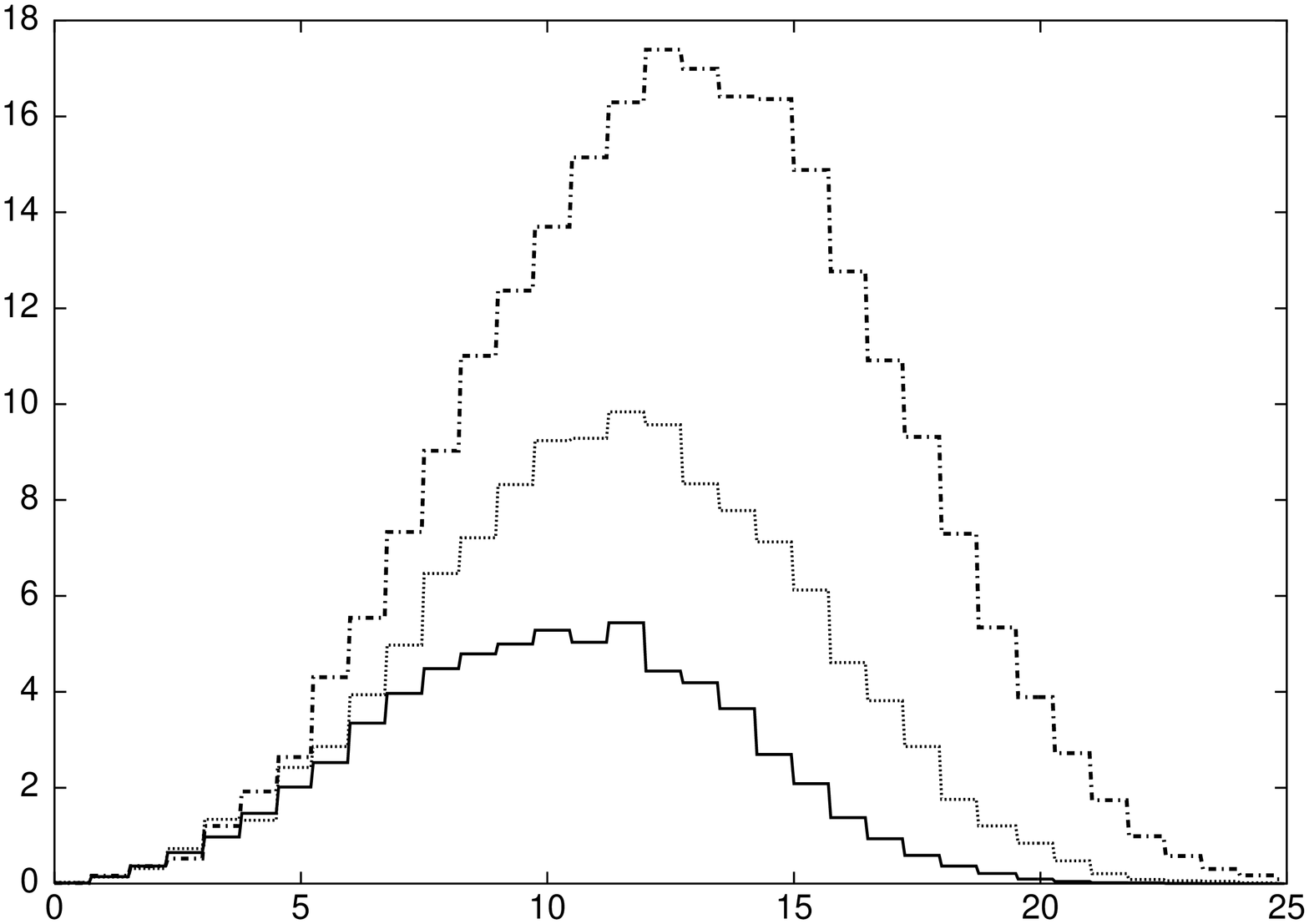,height=10.5cm,width=7.5cm,angle=270} }
 \put(7.2,0.2) {$\xi=\ln(1/x)$}
 \put(1.8,4.5) {{\Large $\frac{\d N_\chi}{\d\xi}$}}
\end{picture}
\caption{\label{LSP_1000}
Neutralino spectra from SUSY QCD Monte Carlo simulation for 
$M_X=10^{12}$~GeV (bottom),
$M_X=10^{13}$~GeV (middle) and
$M_X=10^{14}$~GeV (top), 
all for $\Msusy=1$~TeV.} 
\end{figure}

\begin{figure}
\begin{picture}(15,9)
 \put(2.5,8.3) { 
         \epsfig{file=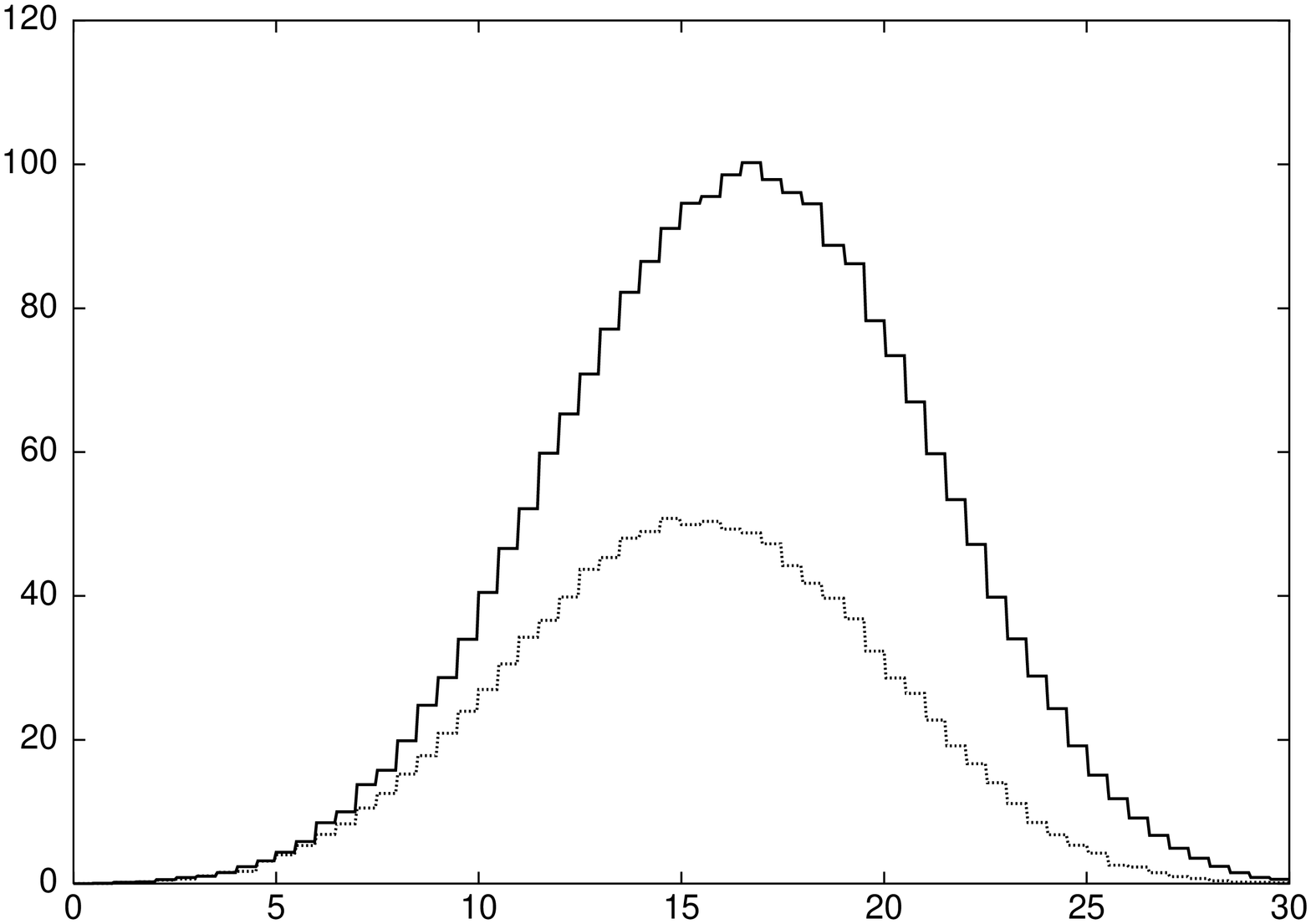,height=10.5cm,width=7.5cm,angle=270} }
 \put(7.2,0.2) {$\xi=\ln(1/x)$}
 \put(1.8,4.5) {{\Large $\frac{\d N_\chi}{\d\xi}$}}
\end{picture}
\caption{\label{LSP_M16}
Neutralino spectra from SUSY QCD Monte Carlo simulation for 
$\Msusy=200$~GeV (top) and $\Msusy=1$~TeV (bottom),
both for $M_X=10^{16}$~GeV.}
\end{figure}

\begin{figure}
\begin{picture}(15,9)
 \put(2.5,8.3) { 
         \epsfig{file=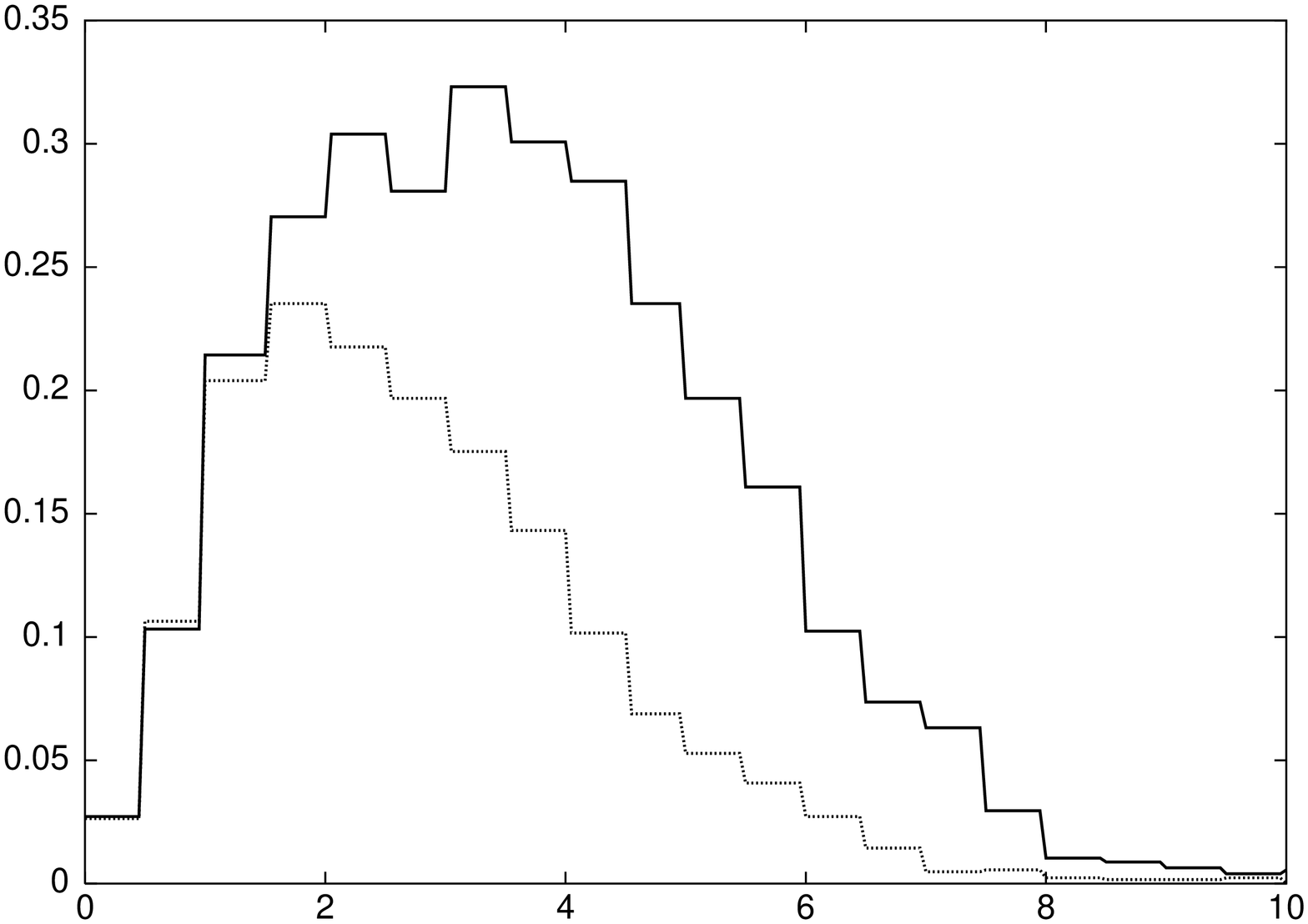,height=10.5cm,width=7.5cm,angle=270} }
 \put(7.2,0.2) {$\xi=\ln(1/x)$}
 \put(1.8,4.5) {{\Large $\frac{\d N_\chi}{\d\xi}$}}
\end{picture}
\caption{\label{LSP_M6}
Neutralino spectra from SUSY QCD Monte Carlo simulation for 
$\Msusy=200$~GeV (top) and $\Msusy=1$~TeV (bottom),
both for $M_X=10^{6}$~GeV.}
\end{figure}

\begin{figure}
\begin{picture}(15,9)
 \put(2.5,8.3) { 
         \epsfig{file=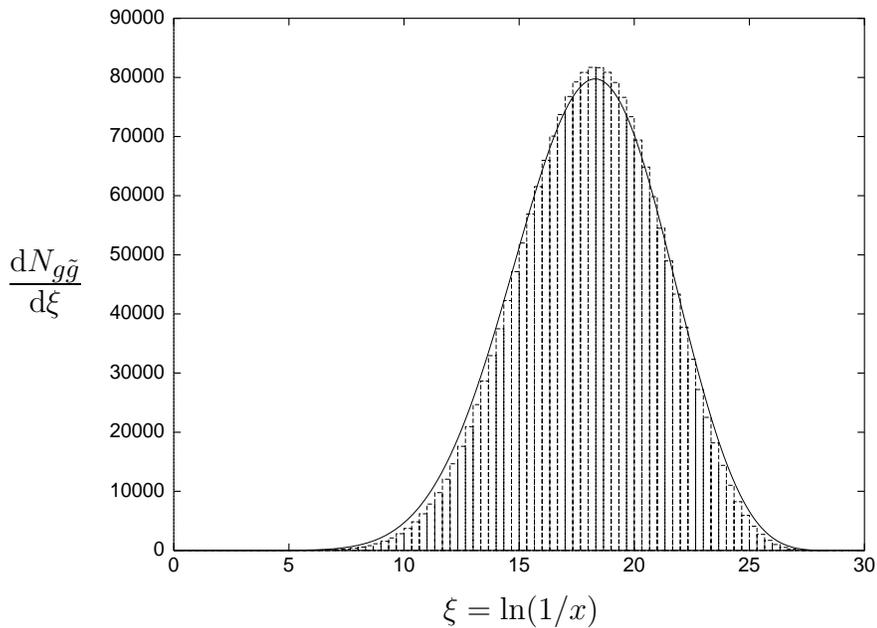,height=10.5cm,width=7.5cm,angle=270} }
 \put(7.2,0.2) {$\xi=\ln(1/x)$}
 \put(1.4,4.5) {{\Large $\frac{\d N_{g\tilde g}}{\d\xi}$}}
\end{picture}
\caption{\label{gg_p_12}
Parton spectrum from SUSY QCD Monte Carlo simulation (boxes) and of 
the SUSY QCD Limiting Spectrum 
(solid line) for $M_X=10^{12}$~GeV. Both for gluons and gluinos only.
The coupling constant in the Monte Carlo simulation is frozen at 
$\tilde{t}< 0.9$~GeV$^2$, and $b_s=3$ is fixed in both cases.}
\end{figure}

\begin{figure}
\begin{picture}(15,9)
 \put(2.5,8.3) { 
          \epsfig{file=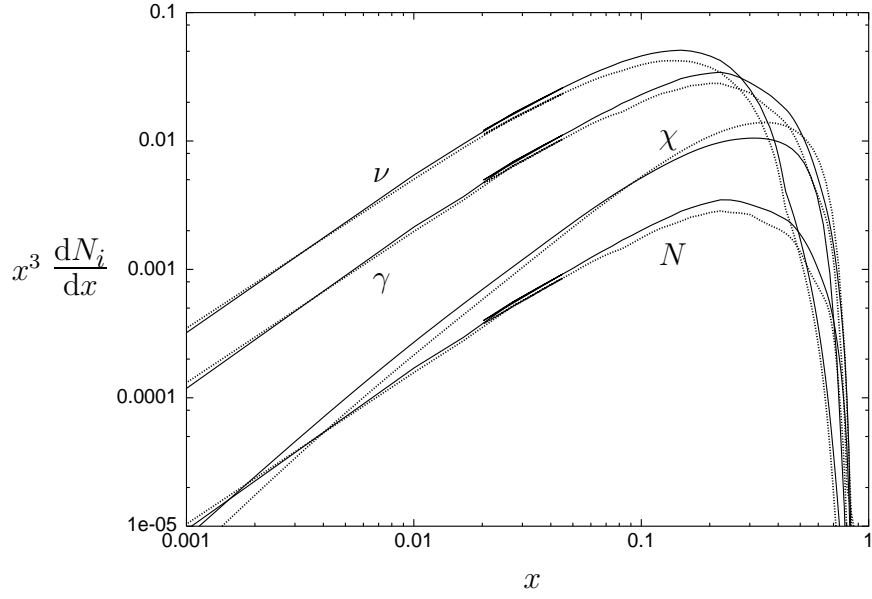,height=10.5cm,width=7.5cm,angle=270} }
 \put(6.2,5.8) {$\nu$}
 \put(6.2,4.4) {$\gamma$}
 \put(10.0,6.3) {$\chi$}
 \put(10.0,4.7) {$N$}
 \put(8.2,0.4) {$x$}
 \put(1.4,4.5) {$x^3\;${\Large $\frac{\d N_i}{\d x}$}}
\end{picture}
\caption{Neutrino, gamma and nucleon fragmentation spectra from
SUSY QCD Monte Carlo simulations for
$M_X=10^{12}$~GeV (solid lines) and $10^{14}$~GeV (dotted lines), all
for $\Msusy=200$~GeV. 
\label{frag}}
\end{figure}

\end{document}